\documentclass[12pt, a4paper]{article}

\usepackage{url}

\usepackage{hyperref}
\usepackage{breakurl}
\usepackage{newtxtext}
\usepackage{latexsym}
\usepackage{graphics}
\usepackage[authoryear]{natbib}
\usepackage{amsmath}
\usepackage{amsfonts}
\usepackage{multirow}
\usepackage{colortbl}
\usepackage{array,ragged2e}
\usepackage{bbm}
\usepackage{enumitem}
\usepackage{arydshln}
\hypersetup{breaklinks=true}
\bibpunct[, ]{(}{)}{;}{a}{,}{,}

\newcolumntype{R}[1]{>{\RaggedLeft}p{#1}}

\renewcommand{\vec}[1]{\boldsymbol{{#1}}}
\providecommand{\myth}[1]{${#1}^{\mathrm{th}}$}

\newcommand{\vtheta}{\vec{\theta}}
\newcommand{\pt}[2]{\vtheta_{{#1}}^{({#2})}}
\newcommand{\vxi}{\vec{\xi}}
\providecommand{\wt}[2]{\omega_{{#1}}^{({#2})}}
\providecommand{\data}[1]{\vec{y}_{{#1}}}
\newcommand{\lik}[2]{p\left(\data{{#1}}\lvert {{#2}}\right)}
\newcommand{\ninterval}[1]{[t_{{#1} - 1}, t_{{#1}})}

\newcommand{\var}[1]{\textrm{var}\left({{#1}}\right)}
\newcommand{\tdens}[2]{\pi^{\textrm{{#1}}}_{{#2}}(\vtheta)}
\newcommand{\mcmcdens}[1]{\tdens{MCMC}{{#1}}}
\newcommand{\smcdens}[1]{\tdens{SMC}{{#1}}}

\newcommand{\conf}[1]{{#1}^{\textrm{conf}}}
\newcommand{\doc}[1]{{#1}^{\textrm{doc}}}

\begin{document}

\title{Efficient real-time monitoring of an emerging influenza epidemic: how feasible?}
\author{\protect\parbox{\textwidth}{\protect\centering Paul
    J. Birrell$^1$, Lorenz Wernsich$^1$,
    Brian D. M. Tom$^1$, Leonhard Held$^3$,\\ Gareth O. Roberts$^4$, Richard G. Pebody$^2$, Daniela De Angelis$^{1,2}$}}
\maketitle
\begin{center}
  \emph{${}^1$MRC Biostatistics Unit, Cambridge Institute of Public Health, Cambridge, UK} \\
  \emph{${}^2$Public Health England, London, UK}\\
  \emph{${}^4$Epidemiology, Biostatistics and Prevention Institute, University of Zurich, Switzerland}\\
  \emph{${}^4$Centre for Research in Statistical Methodology, University of Warwick, Coventry, UK}\\
   e-mail for correspondence: daniela.deangelis@mrc-bsu.cam.ac.uk
\end{center}

\begin{abstract}\noindent
  A prompt public health response to a new epidemic relies on the ability to monitor and predict its evolution in real time as data accumulate.  The 2009 A/H1N1 outbreak in the UK revealed pandemic data as noisy, contaminated, potentially biased, and originating from multiple sources. This seriously challenges the capacity for real-time monitoring. Here we assess the feasibility of real-time inference based on such data by constructing an analytic tool combining an age-stratified SEIR transmission model with various observation models describing the data generation mechanisms. As batches of data become available, a sequential Monte Carlo (SMC) algorithm is developed to synthesise multiple imperfect data streams, iterate epidemic inferences and assess model adequacy amidst a rapidly evolving epidemic environment, substantially reducing computation time in comparison to standard MCMC, to ensure timely delivery of real-time epidemic assessments. In application to simulated data designed to mimic the 2009 A/H1N1 epidemic, SMC is shown to have additional benefits in terms of assessing predictive performance and coping with parameter non-identifiability.
  \bigskip

\end{abstract}

\section{Introduction}\label{sec:intro}

A pandemic influenza outbreak has the potential to place a significant burden upon healthcare systems. The capacity to monitor and predict its evolution as data progressively accumulate, therefore, is a key component of preparedness strategies for a prompt public health response.

Statistical approaches to real-time monitoring have been used for a number of infectious diseases including: prediction of swine fever cases \citep{MeeKJD02}; online estimation of a time-evolving effective reproduction number $R(t)$ for SARS \citep{WalT04, CauBTV06} and for generic emerging disease \citep{BetR08}; inference of the transmission dynamics of avian influenza in the UK poultry industry \citep{JewKCR09}; and forecasting of Ebola \citep{VibSGAFMZCSVR18}.

Typically, however, this work relies on the availability of direct data on the number of new cases of an infectious disease over time. In practice, direct data are seldom available,  as illustrated by the 2009 outbreak of pandemic A/H1N1pdm influenza in the United Kingdom (UK). More likely, multiple sources of data exist, each indirectly informing the epidemic evolution, each subject to possible sources of bias. These data typically come from routine influenza surveillance systems reporting interactions with healthcare services. They are often: biased towards the more severe cases; subject to the changing healthcare-seeking behaviours of the population; contaminated with cases of people experiencing influenza-like illness; and 
heavily influenced by governmental policies.
These features call for more complex modelling, requiring the synthesis of information from a range of data sources in real time.

In this paper we tackle the problem of online inference and prediction in an influenza pandemic in this more realistic situation. We address this starting from the work of \cite{BirKGCPHCZWPD11} who retrospectively reconstructed the A/H1N1 pandemic in a Bayesian framework using multiple data streams collected over the course of the pandemic.
In \cite{BirKGCPHCZWPD11}, posterior distributions of relevant epidemic parameters and related quantities are derived through Markov Chain Monte Carlo (MCMC) methods which, if used in real-time, pose important computational challenges. MCMC is notoriously inefficient for online inference as it requires repeat browsing of the entire data history as new data accrue. This motivates a more efficient algorithm. Potential alternatives include refinements of MCMC \citep[e.g.][]{JewKCR09, BanGLRarX} and Bayesian emulation \citep[e.g][]{FarBCD14}, where the model is replaced by an easily-evaluated approximation readily prepared in advance of the data assimilation process.
Here, we explore Sequential Monte Carlo (SMC) methods \citep{DouJ09}. As batches of data arrive at times $t_1, \ldots, t_K$, SMC techniques allow computationally efficient online inference by combining the posterior distribution $\pi_k(\cdot)$ at time $t_k, k = 0, \ldots, K$ with the incoming batch of data to obtain an estimate for $\pi_{k + 1}(\cdot)$. A further advantage of SMC is that it automatically provides all the posterior predictive distributions necessary to make one-step ahead probabilistic forecasts of the incoming data. In a pandemic context, monitoring the appropriateness of a model is vital to avoid making public health decisions on the basis of mis-specified models. Through formal assessment of the quality of these one-step ahead forecasts \citep{HelMB17}, timely checks of model adequacy and, if necessary, swift adaptations of the model can be made.

Use of SMC in the real time monitoring of an emerging epidemic is not new.   \cite{OngCCLLLTG10}, \cite{DukLP12}, \cite{SkvR12}, \cite{DurKB13}, \cite{CamKAWFBPCGSTEF15} and \cite{FunCKEE18} for instance, provide examples of real time estimation and prediction for deterministic and stochastic models describing the dynamics of influenza and Ebola epidemics. These models, again, only include a single source of information that has either been pre-processed or is free of any sudden or systematic changes. 

In what follows we advance existing literature in three ways: we include a number of data streams, realistically mimicking current data availability in the UK; we consider the situation where a public health intervention introduces a shock to the system, critically disrupting the ability to track the posterior distribution over time; and we demonstrate how the use of SMC can facilitate online assessment of model adequacy.

The paper is organised as follows: in Section \ref{sec:recon} the model in \cite{BirKGCPHCZWPD11} is reviewed focusing on the data available and the computational limitations of the MCMC algorithm in a real time context; in Section \ref{sec:SMC} the idea of SMC is introduced and the algorithm of \cite{GilB01} is described; 
in Section \ref{sec:sim} 
results are presented from the application of Gilks and Berzuini's SMC algorithm to data simulated to mimic the 2009 outbreak and illustrate the challenges posed by the presence of the informative observations induced by system shocks; in Sections \ref{sec:inf.theory} and  \ref{sec:informative} adjusted SMC approaches that address such challenges are assessed; we conclude with Section \ref{sec:discussion} in which the ideas explored in the paper are critically reviewed and outstanding issues discussed.

\section{A model for pandemic reconstruction}\label{sec:recon}
\begin{figure}[!ht]
\centering
  \resizebox{\textwidth}{!}{\includegraphics{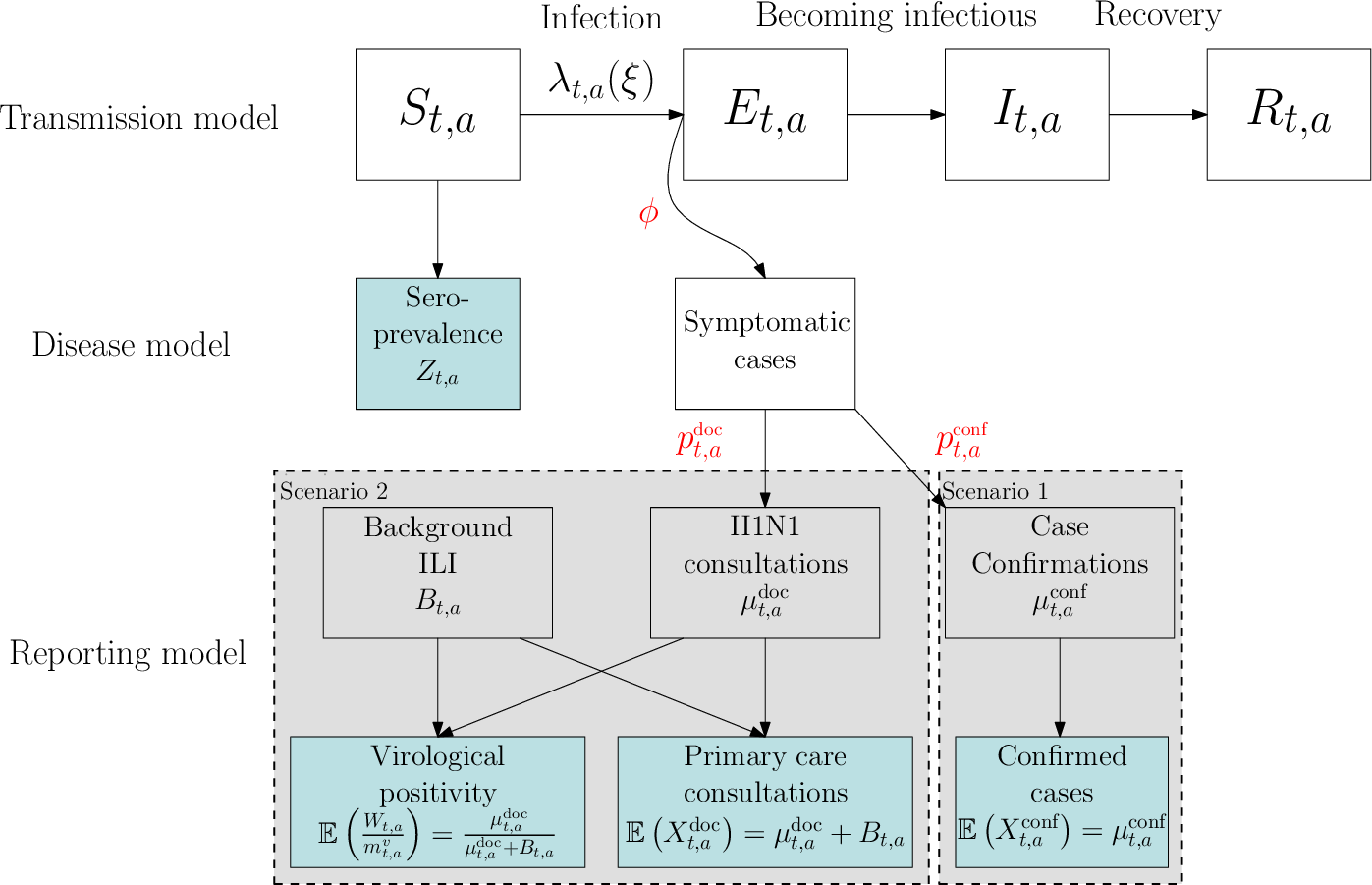}}
\caption{Schematic diagram showing multiple epidemics surveillance sources linking to an SEIR epidemic model via an observation and reporting model. 
The shaded blue boxes represent observed data streams.} \label{fig:link_schema}
\end{figure}

\cite{BirKGCPHCZWPD11} estimate the transmission of a novel influenza virus among a fixed population
stratified into $A$ age groups (see Figure \ref{fig:link_schema}).  Disease transmission is approximated by a deterministic age-structured Susceptible (S), Exposed (E), Infectious (I), Recovered (R) model 
described by a system of differential equations
evaluated at discrete times $t_k = k\delta t, k = 0, \ldots, K$, with $\delta t = 0.5$ days. 
Under this discretisation, the number of new infections in interval $[t_{k-1},t_k)$ is 
\begin{equation}\label{eqn:nni}
\Delta_{t_k, a}=S_{t_{k - 1}, a}\lambda_{t_{k - 1}, a}
\end{equation}
where $\Delta_{t_k, a} \equiv \Delta_{t_k, a}(\vxi)$ for a vector of transmission parameters $\vxi$ and
\begin{equation}\label{eqn:foi}
\lambda_{t_k, a} \equiv \lambda_{t_k, a}(\vxi) = 1 - \prod_{b = 1}^A \left\{\left(1 - \
M_{t_k}^{(a, b)}R_{0}(\psi) / d_I\right)^{{I}_{t_k, b}} \right\}\delta t
\end{equation}
is the time- and age-varying force of infection, the rate at which susceptible individuals become infected. In (\ref{eqn:foi})
$R_{0}(\psi)$ is the basic reproduction number, the expected number of secondary infections caused by a single primary infection in a fully susceptible population, parameterised in terms of the epidemic growth rate $\psi$; $\vec{M}_{t_k}(\vec{m})$ represent time-varying mixing matrices, parameterised by $\vec{m}$, with $M_{t_k}^{(a, b)}(\vec{m})$ giving the relative rates of effective contacts between individuals of each pair of age groups $(a, b)$ at time $t_k$; and $d_L$ and $d_I$ are the mean latent and infectious periods, respectively.
The initial conditions of the system are determined by a further parameter $\nu$. Fixing $d_L = 2$ days, the vector of transmission dynamics parameters is $\vxi = (\psi, \nu, d_I, \boldsymbol{m})$.

There is no direct information to estimate $\vxi$ as the transmission process is unobserved. \cite{BirKGCPHCZWPD11} describe how $\vxi$ can be inferred from the combination of different sources linked to the latent transmission through a number of observational models (see Figure \ref{fig:link_schema}).

A first source of information is provided by a series of cross-sectional serological survey data $Z_{t_k,a}$ on the presence of immunity-conferring antibodies in the general population. Denoting by $N_a$ the population size in age group $a$ and $m^{\textrm{s}}_{t_k, a}$ the number of blood sera samples tested in time interval $\ninterval{k}$, it is assumed that
\begin{equation}
Z_{t_k, a} \sim \textrm{Bin}\left(m^{\textrm{s}}_{t_k, a}, 1 - \frac{S_{t_k, a}}{N_a}\right)\label{eqn:sero.bin}
\end{equation}
informing directly the number of susceptibles $S_{t_k,a} \equiv S_{t_k, a}(\vxi)$ in age group $a$ at the end of the \myth{k} time-step. A second source is the time series of  
virologically confirmed infections ({\it e.g.} admission to intensive care) $\conf{x}_{t_k, a}$ or the number $\doc{x}_{t_k, a}$ of consultations at general practitioners (GP) for influenza like illness (ILI). Data on consultations are contaminated by a ``background'' component of individuals attending GP for non-pandemic ILI, strongly influenced by a public's volatile sensitivity to governmental advice.
Both $\conf{x}_{t_k, a}$ and $\doc{x}_{t_k, a}$ are assumed to be realisations of negative binomial distributions here expressed in a mean-dispersion
$(\mu, \eta)$ parameterisation, such that if $X\sim \textrm{NegBin}(\mu, \eta)$, then $\mathbb{E}(X) = \mu$, $\var{X} = \mu(\eta + 1)$, {\it i.e.}
\begin{equation}\label{eqn:negbin}
\conf{X}_{t_k, a} \sim \textrm{NegBin}\left(\conf{\mu}_{t_k, a}, \eta_{t_k}\right)
\end{equation}
and 
\begin{equation}\label{eqn:negbin.gp}
\doc{X}_{t_k, a} \sim \textrm{NegBin}\left(\doc{\mu}_{t_k, a} + B_{t_k, a}, \eta_{t_k}\right).
\end{equation}
In \eqref{eqn:negbin.gp} the contamination $B_{t_k, a}$ is appropriately parameterised in terms of parameters $\vec{\beta}^B$ (see Section \ref{sec:sim}) and both $\conf{\mu}_{t_k, a}$ and $\doc{\mu}_{t_k, a}$ are expressed through a convolution equation, resulting from the process of becoming infected and experiencing a time delay between infection and the relevant health-care event (see Figure \ref{fig:link_schema}). This convolution for $\doc{\mu}_{t_k, a}$ is
\begin{equation}\label{eqn:convolution}
\doc{\mu}_{t_k, a} = \phi \doc{p}_{t_k,a} \sum_{v = 0}^k \Delta_{t_v,a} f(k-v),
\end{equation}
where the (discretised) delay  probability mass function $f(\cdot)$ accounts for both the time from infection to symptoms and the time from symptoms to GP consultation (see Figure \ref{fig:link_schema}). Note that $\mu^e_{t_k,a} \equiv \mu^{\textrm{e}}_{t_k,a}(\vtheta)$ where $e \in \{\textrm{conf}, \textrm{doc}\}$ and $\vtheta=\{\vxi, \phi, p^{e}_{t_k, a},\eta_{t_k}, \vec{\beta}^B\}$.

The signal $\doc{\mu}_{t_k, a}$ can only be identified by additional virological data from sub-samples of size $m^{\textrm{v}}_{t_k, a}$ of the primary care consultations.
The number of swabs testing positive for the presence of the pandemic strain $W_{t_k, a}$ in each sample is assumed to be distributed:
\begin{equation}
W_{t_k, a} \sim \textrm{Bin}\left(m^{\textrm{v}}_{t_k, a}, 1 - \frac{B_{t_k, a}}{\doc{\mu}_{t_k, a} + B_{t_k, a}}\right).\label{eqn:viro.bin}
\end{equation}

\subsection{Inference} \label{sec:MCMC}
To estimate $\vtheta$, 
\cite{BirKGCPHCZWPD11} develop a Bayesian approach and use a Markov Chain-Monte Carlo (MCMC) algorithm to derive the posterior distribution of $\vtheta$ on the basis of $245$  days of primary care consultation and swab positivity data, confirmed case and cross-sectional serological data.
Their MCMC algorithm is a naively adaptive random walk Metropolis algorithm, requiring $7\times10^5$ iterations, requiring in excess of $6.3\times10^6$ evaluations of the transmission model and/or convolutions of the kind in equation \eqref{eqn:convolution}. MCMC is not easily adapted for parallelised computation, but the likelihood calculations allow for some small-scale parallelisation. The MCMC were thus optimally run on a desktop computer with 8 parallel 3.6GHz Intel(R) Core(TM) i7-4790 processors, requiring run-times of almost four hours. 
Although this runtime might not be prohibitive for real-time inference, this implementation leaves little margin to consider multiple code runs or alternative model formulations.
In a future pandemic there will be a greater wealth of data facilitating a greater degree of stratification of the population \citep{SPIM11}. With increasing model complexity comes rapidly increasing MCMC run-times, which can be efficiently addressed through use of SMC methods.

\section{An SMC alternative to MCMC}\label{sec:SMC}

Let $\vec{Y}_t$ denote the vector of all random quantities in \eqref{eqn:sero.bin}-\eqref{eqn:viro.bin}, and let $\data{t}$ be the observed values of $\vec{Y}_t$. Online inference involves the sequential estimation of posterior distributions 
$\pi_k(\vec{\theta}) = p({\vec\theta}\lvert\data{1:k}) \propto \pi_0(\vec{\theta})p(\data{1:k}\vert \vec{\theta}), \; k = 1, \ldots, K$,
where $\pi_0(\vec{\theta})$ indicates the prior for $\vec{\theta}$.
Estimation of any epidemic feature, {\it e.g.} the assessment of the current state of the epidemic or prediction of its future course, follows from estimating $\vtheta$.

Suppose at time $t_k$ a set of $n_k$ particles $\{\vtheta_k^{(1)},\ldots,\vtheta_k^{(n_k)}\}$, with associated weights 
$\{\wt{k}{1},\ldots\allowbreak,\wt{k}{n_k}\}$, approximate a sample from the target distribution $\pi_k(\cdot)$. On the arrival of the next batch of data $\data{k+1}$, $\pi_k(\cdot)$ is used as an importance sampling distribution to sample from $\pi_{k + 1}(\cdot)$. In practice, this involves a reweighting of the particle set. 
The particles are reweighted according to the importance ratio, $\pi_{k+1}(\cdot)/\pi_{k}(\cdot)$, which reduces to the likelihood of the incoming data batch, i.e:
\begin{equation}\label{eqn:alg.reweight}
\wt{k + 1}{j} \propto \wt{k}{j} \frac{\pi_{k + 1}\left(\vtheta_k^{(j)}\right)}{\pi_k\left(\vtheta_k^{(j)}\right)} = \wt{k}{j}\lik{k + 1}{\vtheta_{k}^{(j)}}.
\end{equation}
Eventually, many particles will carry relatively low weight, leading to sample degeneracy as progressively fewer particles contribute meaningfully to the estimation of $\pi_k(\cdot)$.
A measure of this degeneracy is the effective sample size (ESS) \citep{LiuC95},
\begin{equation}\label{eqn:ESS}
\textrm{ESS}\left(\left\{\wt{k}{\cdot}\right\}\right) = \frac{\left(\sum_{j = 1}^{n_k}\wt{k}{j}\right)^2}{\sum_{j = 1}^{n_k} {\wt{k}{j}}^2}.
\end{equation}
The ESS is the ``required size of an independent sample drawn directly from the target distribution to achieve the same estimating precision attained by the sample contained in the particle set'' \citep{CarCF99}, and as such, values of the ESS that are small in comparison to $n_k$ are indicative of an impoverished sample.

This degeneracy can be tackled in different ways. \cite{GorSS93} introduced a resampling step, removing low weight particles and 
jittering the remainder. This jittering step was formalised by \cite{GilB01} using Metropolis-Hastings (MH) steps to rejuvenate the sample. \cite{Fea02} and \cite{Cho02} provide more general treatises of this SMC method, with \cite{Cho02} labelling the algorithm `iterated batch importance sampling'. This was extended by \cite{DelDJ06} who unify the static estimation of $\vtheta$ with the filtering problem (estimation of a state vector, $\vec{x}_k$).

Here we adapt the resample-move algorithm of \cite{GilB01}, investigating its real-time efficiency in comparison to successive use of MCMC. The MH steps rejuvenating the sample constitute the computational bottle-neck in resample-move as they require a browsing of the whole data history to evaluate the full likelihood, not just the most recent batch. For fast inference, the number of such steps should be minimised, without risking Monte Carlo error through sample degeneracy. The resulting algorithm is laid out in full below. It is presumed that it is straightforward to sample from the prior distribution $\pi_0(\vtheta)$.

\subsection{The algorithm}\label{sec:alg}

\begin{enumerate}
\item {{\bf Set} $k = 0$. Draw a sample $\{\vtheta_0^{(1)},\ldots,\vtheta_0^{(n_0)}\}$ from the prior distribution, $\pi_0(\vtheta)$, set the weights $\wt{0}{j} = 1 / n_0, \forall j$.}
\item {{\bf Set}\label{pt:init} $k = k + 1$. Observe a new batch of data $\vec{Y}_{k} = \data{k}$. Reweigh the particles so that the \myth{j} particle has weight, 
$\tilde\omega_{k}^{(j)} \propto \wt{k - 1}{j} p\left(\data{k} \lvert \vtheta_{k - 1}^{(j)}\right)$.}
\item {{\bf Calculate the effective sample size}. Set $\omega^{*(j)}_{k} = \tilde\omega_{k}^{(j)} / \sum_i \tilde\omega_{k}^{(i)}, \forall j$. If $\textrm{ESS}\left(\left\{\omega_{k}^{*(\cdot)}\right\}\right) > \epsilon_L n_{k - 1}$ set $\vtheta_k^{(j)} = \vtheta_{k - 1}^{(j)}$, $\wt{k}{j} = \omega^{*(j)}_{k}$, $n_k = n_{k - 1}$ and return to point (\ref{pt:init}), else go next}.

\item {{\bf Resample}. Choose $n_k$ and sample  $\{\tilde\vtheta_{k}^{(j)}\}_{j = 1}^{n_k}$
from the set of particles $\{\vtheta_{k - 1}^{(j)}\}_{j = 1}^{n_{k - 1}}$ with corresponding probabilities $\{\omega_{k}^{*(j)}\}_{j = 1}^{n_{k - 1}}$. Here, we have used residual resampling \citep{LiuC98}. Re-set $\omega_{k}^{(j)} = 1 / n_{k}$. \label{pt:Rejuvenation}
}
\item {{\bf Move}: For each $j$, move from $\tilde{\vtheta}_{k}^{(j)}$ to  $\vtheta_{k}^{(j)}$ via a MH kernel $\mathcal{K}_{k}\left(\tilde{\vtheta}_{k}^{(j)}, \vtheta_{k}^{(j)};\gamma\right)$.  If $k < K$, return to point (\ref{pt:init}).\label{pt:move}}
\item {{\bf End}: $\left\{\left(\wt{K}{1},\pt{K}{1}\right), \ldots, \left(\wt{K}{n_K},\pt{K}{n_K}\right)\right\}$ is a weighed sample from $\pi_K(\cdot)$.}
\end{enumerate}

There are a number of algorithmic choices to be made, including tuning any parameters, $\gamma$, of the MH kernel and the rejuvenation threshold, $\epsilon_L$. In a real-time setting, it may not be possible  to tune an algorithm ``on the fly'', so the system has to work ``out of the box'', either through prior tuning or through being adaptive \citep{FeaT13}. In what follows we set $\epsilon_L=0.5$ \citep{JasSDT11} and we
focus on the key factors affecting  the performance of the algorithm in real-time, {\it i.e.} the MH kernel.

\subsubsection{Kernel choice}\label{sec:ker.choice}

\paragraph{Correlated random walk}
A correlated random walk proposes values in the neighbourhood of the current particle: 
\begin{equation}
\vtheta^* \lvert \tilde\vtheta^{(j)}_k \sim \mathrm{N}\left(\tilde\vtheta^{(j)}_k, \gamma \bar{\vec{\Sigma}}_k\right)\label{eqn:appCorr},
\end{equation}
where $\bar{\vec{\Sigma}}_k$ is the sample variance-covariance matrix of the weighted sample $\{\tilde\omega_{k}^{(\cdot)} \cdot \vtheta_{k-1}^{(\cdot)}\}$.
The advantages here are that the parameter $\gamma$ can be tuned {\it a priori} to guarantee a reasonable acceptance rate, or asymptotic results for the optimal scaling of covariance matrices \citep{RobR01, SheFR10} could be used. Also, the localised nature of these moves should keep acceptance rates high, leading to quick restoration of the value of the ESS. 

\paragraph{Approximate Gibbs'}
An independence sampler that proposes \citep{Cho02}:
\begin{equation}
\vec{\theta^*} \lvert \tilde\vtheta^{(j)}_k \sim \mathrm{N}\left(\bar{\vtheta}_k, \bar{\vec{\Sigma}}_k\right)\label{eqn:appGibbs}
\end{equation}
where $\bar\vtheta_k$ is the sample mean of the $\{\tilde\omega_{k}^{(\cdot)}\cdot \vtheta_{k-1}^{(\cdot)}\}$.
Here, proposals are drawn from a distribution chosen to approximate the target distribution, only weakly-dependent on the current position of the particle.
An accept-reject step is still required to correct for this approximation
. The quality of the approximation depends on $\pi_{k - 1}(\cdot)$ being well represented by the current particle set, there being sufficient richness in the particle weights after the reweighting step and the target density being sufficiently near-Gaussian. Assuming that the multivariate normal approximation to the target is adequate (and it should be increasingly so as more data are acquired) this type of proposal allows for more rapid exploration of the sample space. 

For each type of kernel, both block 
and component-wise (where individual or sub-groups of parameter components are proposed in turn) proposals that use the appropriate conditional distributions derived from \eqref{eqn:appCorr} and \eqref{eqn:appGibbs} are considered. However, the kernels considered in Step \ref{pt:move} of the resample-move algorithm consist of only a single block proposal or a single proposal for each parameter component.

\section{A simulated epidemic}\label{sec:sim}

The suitability of the SMC algorithm for real-time epidemic inference is evaluated against the MCMC algorithm used in \cite{BirKGCPHCZWPD11}, which is taken as a gold-standard. Comparisons are made through application to data simulated from the epidemic model in Figure \ref{fig:link_schema}. The simulation conditions were chosen so that the resulting epidemic would mimic the timing and dynamics of the 2009 A/H1N1 pandemic in England. This epidemic 
was characterised by two distinct waves of infection with a first peak induced by an over-summer school holiday and a second peak occurring during the traditional winter flu season.

\begin{figure}
\centering
\begin{tabular}{@{}l@{}r@{}}
\resizebox{0.5\linewidth}{!}{\includegraphics{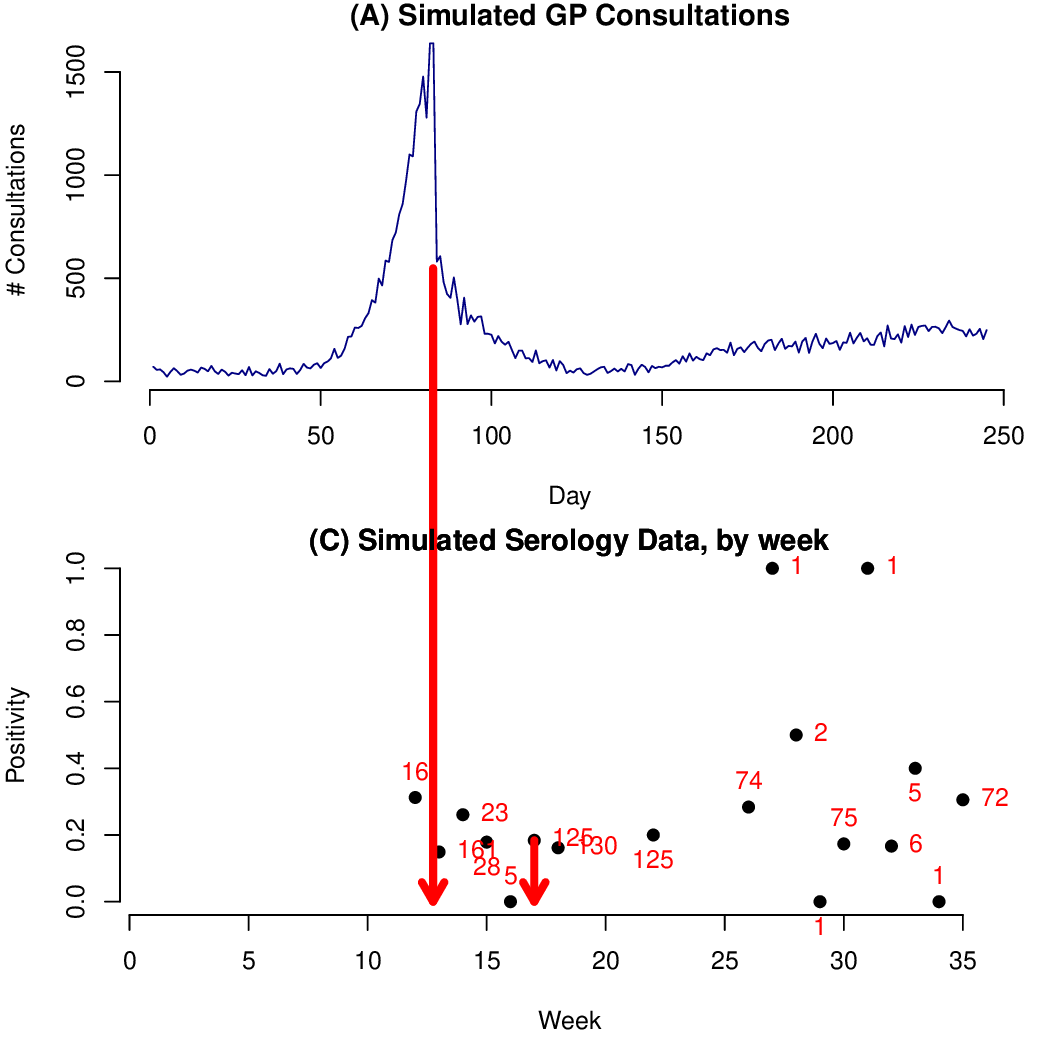}}&
\resizebox{0.5\linewidth}{!}{\includegraphics{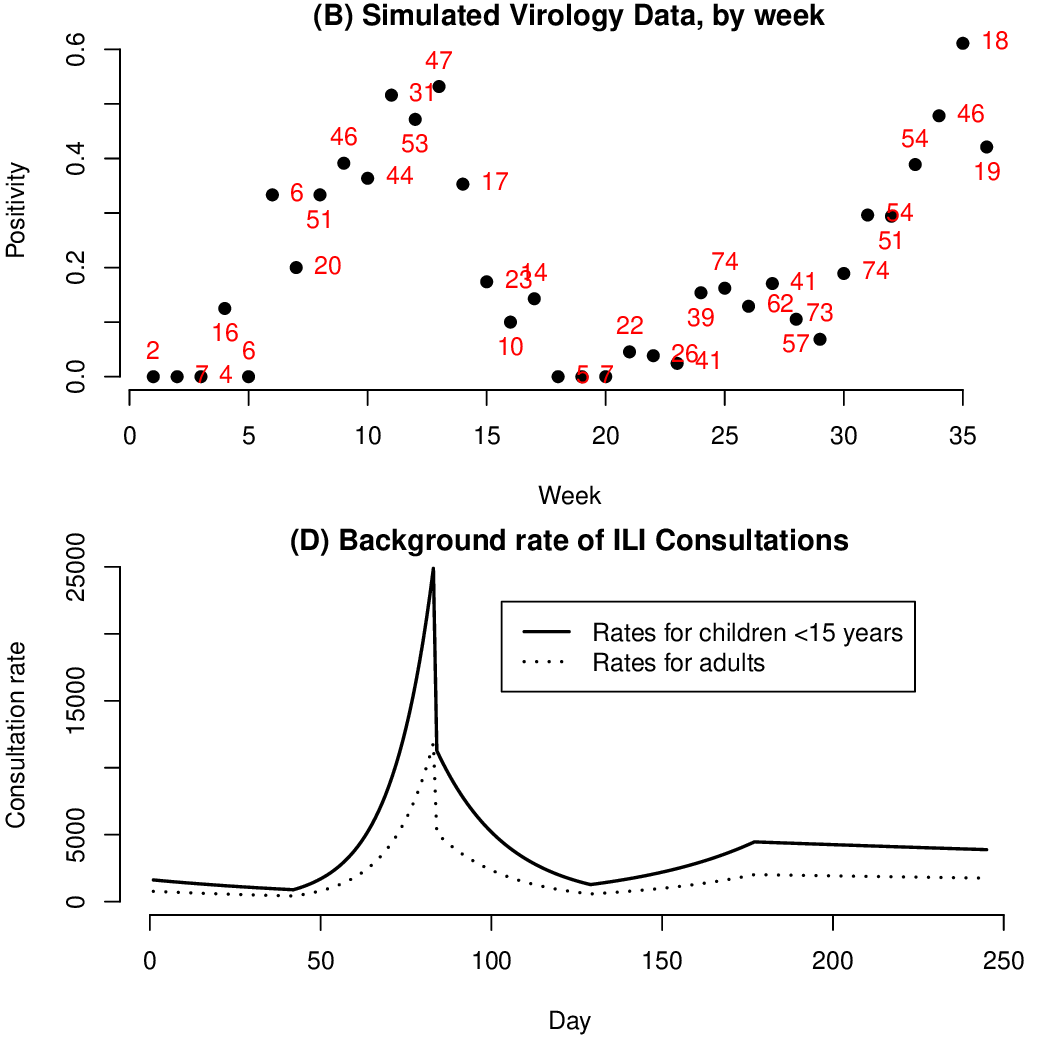}}\\
\end{tabular}
\caption{Top row: (A) Number of doctor consultations $\doc{X}_{t_k, a}$; (B) swab positivity data ($W_{t_k, a}$) with numbers representing the size of the weekly denominator. Bottom row: (C) serological data ($Z_{t_k, a}$); (D) pattern of background consultation rates by age. Arrows between (A) and (C) highlight the timing of some key, informative observations.}\label{fig:gp.data}
\end{figure}

We consider two scenarios (see Figure \ref{fig:link_schema}). In the first, direct information on confirmed cases (e.g. hospitalisation, ICU admissions) is available; in the second we observe the noisy ILI consultations (Equation \eqref{eqn:negbin.gp}). Alongside either of these data, serological data (Equation \eqref{eqn:sero.bin}) are available and, in the second scenario, there are also virological data taken from a sub-sample of the ILI consultations (see Equation \eqref{eqn:viro.bin}). Both scenarios use observations made on 245 consecutive days on a population divided into $A = 7$ age groups, and are characterised by the same underlying epidemic curve, so that the confirmed case and primary care consultation data are subject to similar trends. For both scenarios we introduce a shock at $t_k = 83$ days, similarly to the 2009 pandemic, where a public health intervention is assumed to change the way the confirmed cases or consultations occur and are reported. The simulated data for the second scenario are presented in Figure \ref{fig:gp.data}(A)-(C), where the timing of the shock is indicated by the red arrow linking (A) and (C). 
Table 
A1
in the online appendix presents the model parameters 
together with the values used for simulation. 
Note that the proposed intervention impacts on three groups of parameters by introducing a changepoint: the dispersion in the count data, $\eta$, the proportion of infections that appear in the data $p^{\cdot}$,
and in Scenario 2 the age-specific (i.e. child and adult specific) background consultation rates, $B_{t_k,a}$, which develop over time according to a log-linear spline with a discontinuity at $t_k = 83$. The spline 
is plotted, by age group, in Figure \ref{fig:gp.data}(D) 
and its parameterisation as a function of the 9-dimensional parameter $\vec{\beta}^B$ is given in Section A1 
of the online appendix. 

Real-time monitoring of the epidemic will begin after an initial outbreak stage, taken here to be the first $50$ days. An MCMC implementation of the model is carried out at times $t_k = 50, 70, 83, 120, 164$ and 245 days and the SMC algorithm is then used to propagate the MCMC-obtained posteriors over the intervals defined by these timepoints.
For example, the MCMC-obtained estimate $\mcmcdens{50}$ of $\pi_{50}(\vtheta)$ will be used as the initial particle set for the SMC algorithm over the interval 50-70 days. This gives an estimate, $\smcdens{70\lvert 50}$, for $\tdens{}{70}$, which is then compared to $\mcmcdens{70}$. 
The similarity between the two distributions is measured by the K\"ullback-Leibler (KL) divergence of $\smcdens{t_k\lvert\cdot}$  from the ``gold-standard'' reference distribution $\mcmcdens{t_k}$, calculated using multivariate normal approximations to both distributions.

\subsection{{\color{black}Results from a resample-move SMC algorithm}}\label{sec:analysis}

\begin{figure}
\centering
(A) $t_k = 70$\\
\resizebox{0.9\linewidth}{!}{\includegraphics{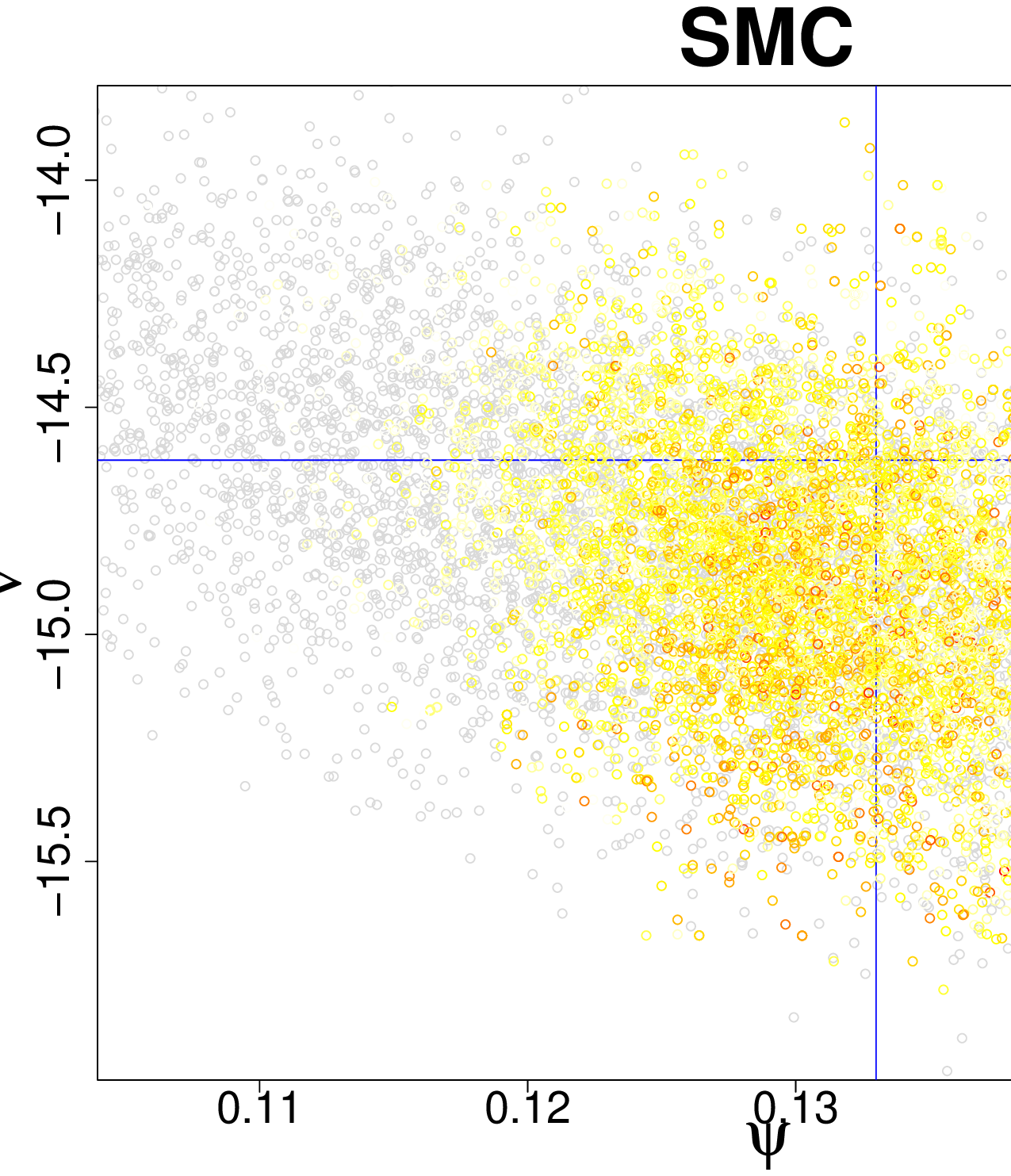}}\\
(B) $t_k = 120$\\
\resizebox{0.9\linewidth}{!}{\includegraphics{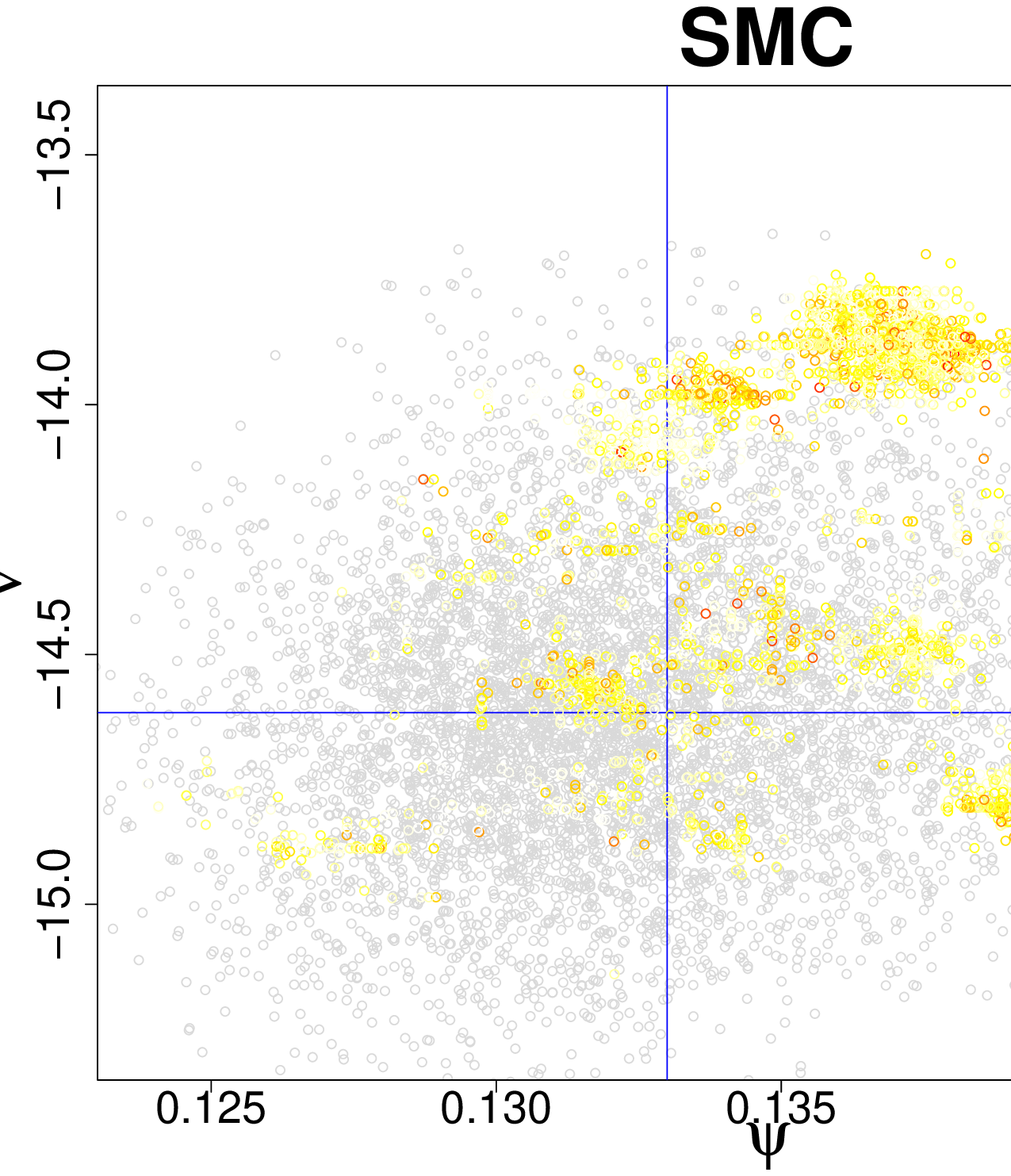}}\\
(C) $t_k = 245$\\
\resizebox{0.9\linewidth}{!}{\includegraphics{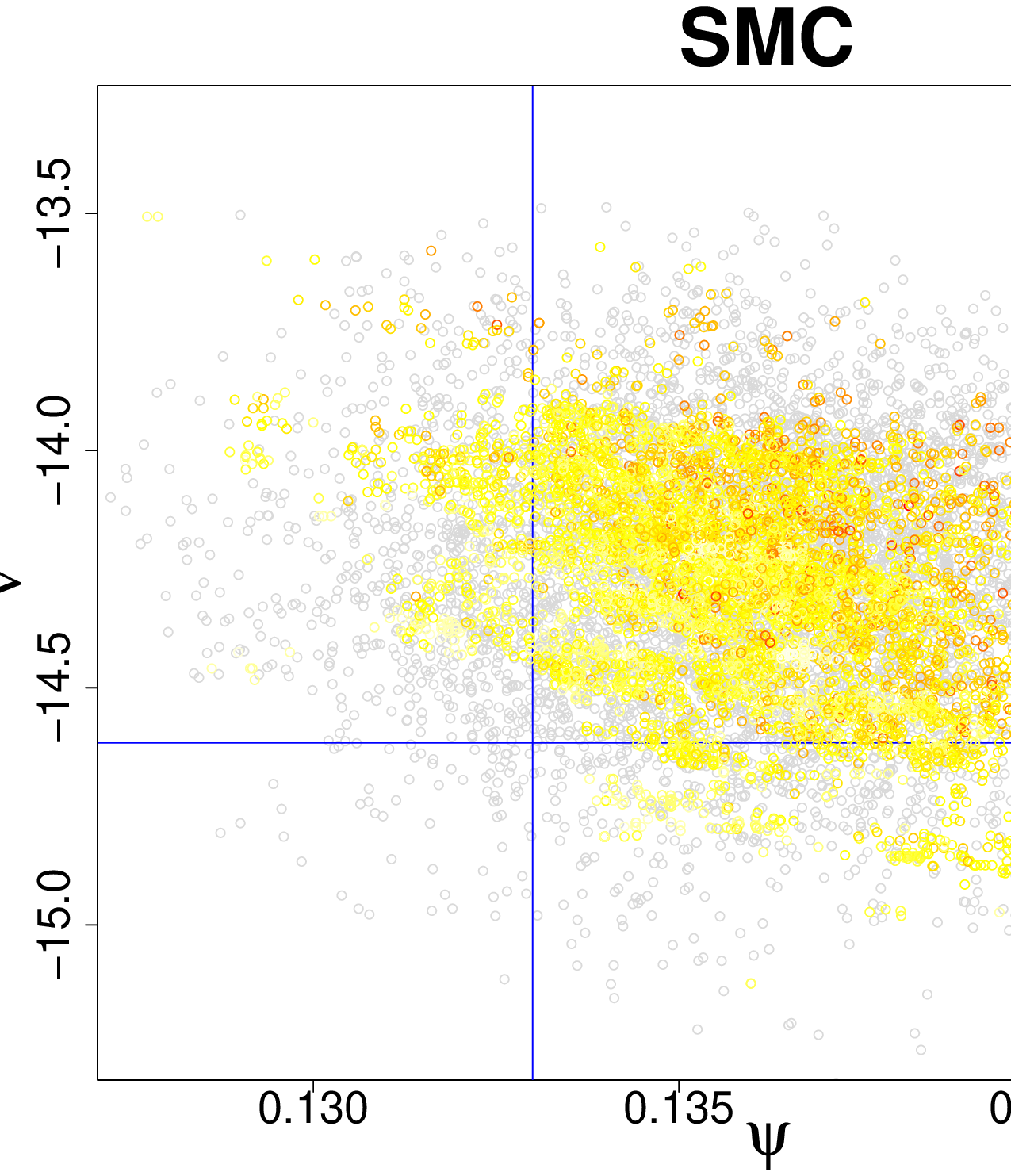}}\\
\caption{\label{fig:ind.chopin.plots}Comparison of SMC-obtained posteriors and MCMC-obtained posteriors at $t_k$ = 70 (A), $t_k$ = 120 (B) and $t_k$ = 245 (C) days, via scatter plots for the parameters $\psi$ and $\nu$. The grey points in both the left and the right panels represent the MCMC-obtained sample at the beginning of the interval, with the overlaid coloured points representing the SMC or MCMC-obtained samples at the end of the interval. In the SMC-obtained samples, the colour of the plotted points represents the weight attached to the particle, with the red particles being those of heaviest weight.}
\end{figure}

In addition to KL, Table \ref{tbl:KL} reports Hellinger and Wasserstein divergences for the posterior distributions from 
Scenario 1, obtained using each of the three different proposal kernels described in Section \ref{sec:ker.choice}. The use of the three divergences ensures that inference is not being unduly influenced by the particular characteristics of any single chosen metric.  
The correlated random-walk \eqref{eqn:appCorr} has the highest KL over the intervals up to 120 days. Beyond 120 days, the divergence between distributions $\pi_k$ and $\pi_{k + 1}$ is small and the random-walk proposals become progressively more able to bridge the gap. The component-wise approximate Gibbs scheme \eqref{eqn:appGibbs} generally outperforms the block updates. 
Figure \ref{fig:ind.chopin.plots} illustrates the performance of the approximate Gibbs component-wise proposal kernel
comparing the SMC-  and MCMC-obtained scatterplots for the parameter components $\psi$ and $\nu$ at $t_k = 70$ (A), $t_k = 120$ (B) and $t_k = 245$ (C). There is close correspondence  between the SMC and MCMC obtained distributions at $t_k = 70$ and $t_k = 245$, but substantial departure at $t_k = 120$. This is the only interval for which the block updates perform better (in terms of divergence, Table \ref{tbl:KL}). All of the above findings are consistent irrespective of the metric used. As a result, for ease of presentation we will work with the more familiar KL only from here on. Similar phenomena are observed for Scenario 2, but with magnified KL discrepancies due to the increase in dimensionality (see Table 
B2,
online appendix).

\begin{table*}
\caption{\label{tbl:KL}Scenario 1: K\"ullback-Leibler (KL), Hellinger and Wasserstein statistics and likelihood evaluations per day (`Run Time') for each resample-move algorithm. Bootstrap standard errors are given in brackets.}
\centering
\begin{tabular}{clrrr}
\hline
\multirow{2}{*}{\em Intervals}&\multicolumn{1}{c}{\em Proposal}&\multicolumn{1}{c}{\em Correlated}&\multicolumn{1}{c}{\em Component-wise}&\multicolumn{1}{c}{\em Block}\\
&\multicolumn{1}{c}{\em Method}&\multicolumn{1}{c}{\em Random-Walk}&\multicolumn{1}{c}{\em approx. Gibbs}&\multicolumn{1}{c}{\em approx. Gibbs}\\
\hline
\multirow{4}{*}{0-50}&KL&2.83 (0.018)&2.58 (0.011)&2.61 (0.011)\\
&Hellinger&0.852 (0.0012)&0.833 (0.0010)&0.835 (0.00091)\\
&Wasserstein&19700 (670)&12700 (280)&12300 (220)\\
&Run Time&18200&16800&8000\\ \hdashline
\multirow{4}{*}{51-70}&KL&2.00 (0.016)&0.908 (0.013)&1.32 (0.018)\\
&Hellinger&0.768 (0.0021)&0.546 (0.0032)&0.643 (0.0032)\\
&Wasserstein&1710 (57)&112 (2.5)&230 (3.7)\\
&Run Time&21000&21000&8000\\ \hdashline
\multirow{4}{*}{71-83}&KL&4.44 (0.12)&0.929 (0.037)&1.60 (0.037)\\
&Hellinger&0.804 (0.0033)&0.404 (0.0063)&0.513 (0.0042)\\
&Wasserstein&409 (14)&0.936 (0.065)&1.35 (0.077)\\
&Run Time&26923&26923&7692\\ \hdashline
\multirow{4}{*}{84-120}&KL&16.3 (0.39)&6.58 (0.19)&2.09 (0.085)\\
&Hellinger&0.955 (0.0012)&0.865 (0.0026)&0.497 (0.0055)\\
&Wasserstein&10.5 (0.27)&8.66 (0.20)&0.249 (0.0075)\\
&Run Time&20811&17027&10000\\ \hdashline
\multirow{4}{*}{121-164}&KL&0.106 (0.010)&0.113 (0.0086)&0.122 (0.0077)\\
&Hellinger&0.165 (0.0081)&0.169 (0.0067)&0.172 (0.0051)\\
&Wasserstein&0.0342 (0.0045)&0.0441 (0.0049)&0.0355 (0.0049)\\
&Run Time&3182&3182&4773\\ \hdashline
\multirow{4}{*}{165-245}&KL&0.339 (0.013)&0.471 (0.025)&1.15 (0.035)\\
&Hellinger&0.274 (0.0047)&0.296 (0.0065)&0.424 (0.0046)\\
&Wasserstein&0.0976 (0.0097)&0.0406 (0.0044)&0.109 (0.0046)\\
&Run Time&8642&9506&9136\\
\hline
\end{tabular}
\end{table*}

Irrespective of the kernel chosen, it is clear that the basic resample-move SMC algorithm cannot handle the `shock' in the count data occurring at $t_k=83$, which leads to step changes in some model parameters. 
The marginal posterior distributions for the new parameter components move rapidly from day 84 as probability density shifts away from uninformative prior distributions. 
For Scenario 1, the 84-120 day interval is the only one over which the block-update approximate Gibbs method gives the best performance (see KL divergence in Table \ref{tbl:KL}). 
This arises due to the low acceptance of the single full block proposals, ensuring that the ESS remains below $\epsilon_L n_k$ and leading to further rejuvenations at each following time. This frequent rejuvenation better enables the tracking of the shifting posterior distributions over time (slightly reducing the advantage of this algorithm in terms of computation time, Tables \ref{tbl:KL}). Alternatively, component-wise updates lead to a set of nearly unique particles with $\textrm{ESS} \approx n_k$, and fewer subsequent rejuvenations.
 However, even with the block updates, good correspondence between the SMC- and MCMC-obtained posteriors is not achieved after the shock in Scenario 1 until $t_k\approx 100$, and not at all in Scenario 2.

From these initial results it is clear that a modified algorithmic formulation is needed for computationally efficient inference when target posteriors are highly non-Gaussian and/or are moving fast between successive batches of data as a consequence of highly informative observations.

\section{Extending the algorithm - handling informative observations}\label{sec:inf.theory}

A key feature of any improved SMC algorithm must be that the ESS retains its interpretation 
given in Section \ref{sec:SMC}. 
For example, as a proposal scaling for a random-walk proposal tends to zero, ({\it i.e.} $\gamma \downarrow 0$ in Equation \eqref{eqn:appCorr}), acceptance rates will be close to unity, resulting in a set of mostly unique particles and a high value for the ESS. However, in cases where there has been a loss of particle diversity at the resampling stage (because many particles are sampled numerous times) this would give a highly clustered posterior sample, barely distinguishable from the set of resampled particles and definitely not as informative as an independent sample of size $n_k$. Here, 
the ESS, as calculated from the particle weights, is no longer a reliable guide to the quality of the sample.

We look at three possible improvements to the resample-move algorithm of Section \ref{sec:SMC}, to produce an information-adjusted (IA) SMC algorithm that safeguards the ESS as a good measure of the quality of the sample: we address the timing of rejuvenations; we reconsider the choice of kernels used in the rejuvenations; and we address the problem of choosing the number of iterations we need to run the MCMC sampler before the sample is fully rejuvenated.

\subsection{Timing the rejuvenations: a continuous-time formulation}
If there is large divergence between consecutive target distributions $\pi_k$ and $\pi_{k + 1}$, the estimation of intermediate distributions will allow the particle set to move gradually between the two targets \citep{DelDJ06}. These intermediate distributions are generated via tempering \citep{Nea96}, introducing gradually the new batch of data into the likelihood at a range of `temperatures', $\tau \in [0, 1]$.
These distributions are denoted $\pi_{k, \tau}(\vtheta) \propto \pi_{k}(\vtheta) \left\{\lik{k + 1}{\vtheta}\right\}^\tau$.

We choose to think of data $\data{k+1}$ arriving uniformly over the
\myth{(k+1)} 
interval 
and denote $\wt{k + \tau, \tau_0}{j}$ to be the weight attached to a particle
at an intermediate time $t_{k + \tau}$ when the previous rejuvenation took place at time $t_{k + \tau_0}$, with $\tau_0 = 0$ corresponding to no prior rejuvenation within the interval $(t_k, t_{k + 1}]$. Then, for $0 \leq \tau_0 \leq \tau \leq 1$ and indicator function $\mathbbm{1}_A$,
\begin{equation*}
  \tilde\omega_{k + \tau, \tau_0}^{(j)} = \left(\wt{k}{j} + \left(1 - \wt{k}{j}\right)\mathbbm{1}_{\tau_0 > 0}\right)\lik{k+1}{\vtheta^{(j)}}^{\tau - \tau_0}.
\end{equation*}
Therefore, if $ESS(\{\tilde\omega_{k + 1, \tau_0}^{(j)}\}_{j = 1}^{n_k}) < \epsilon_L n_k$ a further rejuvenation would be proposed at time $\tau^*$, such that $\tau^* = \arg \min_{\tau \in (\tau_0, 1)} \{ESS(\tilde\omega^{(j)}_{k + \tau, \tau_0}) - \epsilon_L n_k\}^2$.

\subsection{Choosing kernels - hybrid algorithms}\label{sec:kernels}
As discussed in Section 
\ref{sec:analysis}, each of the possible MH kernels has its own distinct strengths. 
These can be exploited by using a combination of kernels. Full block approximate-Gibbs updates are efficient at reducing the clustering that forms around resampled particles.
Adding a random walk step would allow the proposal of values outside the space spanned by the principal components of $\bar{\vec{\Sigma}}_k$, something of particular necessity if the ESS is very small and $\bar{\vec{\Sigma}}_k$ is close to singularity.

This motivates a hybridisation of the proposal mechanism, done either by using mixture proposals, {\it e.g.} a mixture between the approximate Gibbs' proposals and full block ordinary random walk Metropolis proposals \citep{KanBJ14}, or, as will be used in the remainder, by augmenting full block approximate Gibbs updates with componentwise random walk proposals.

\subsection{How many MH iterations? Multiple proposals and intra-class correlation}\label{sec:ICC}

In the MH-step of the algorithm, there are effectively $n_k$ parallel MCMC chains. Making proposals until all chains have attained convergence would be an inefficiency. The distribution governing the starting states of these chains forms a biased sample from the target distribution obtained through sampling importance resampling \citep{Cho02}. It then seems a reasonable requirement that we carry out MH steps until the chains have collectively `forgotten' their starting values. This can be monitored through an estimate of an intra-class correlation coefficient (ICC), $\rho$. 
Firstly, the particle set is divided into $I$ clusters, each of size $d_i,\; i = 1, \ldots, I$, defined by the parent particle at the resampling stage. For example, if a particular particle is resampled 5 times, it defines a cluster in the new sample with $d_i = 5$. The analysis of variance intra-class correlation coefficient, $r_A$ \citep{DonK80, SokR81}, is used to estimate $\rho$. This estimate is dependent on the mean squared error in a univariate summary statistic, $g_{ij} = g(\vtheta_{ij})$, 
calculated for the \myth{j} particle in the \myth{i} cluster, $\vtheta_{ij}$ both within and between clusters. Here, we choose the `attack rate' of the epidemic, the cumulative number of infections caused by the epidemic:
\begin{equation}\label{eqn:gfunc}
g(\vtheta) = \frac{\sum_{t = 1}^{\infty}\sum_{a = 1}^A\Delta_{t, a}(\vtheta)}{\sum_{a = 1}^A N_a}.
\end{equation}
Details of the calculation of $r_A$ are in Section C of the online appendix.

Prior to the MH-phase of the algorithm there is no within-class variation and $r_A = 1$. However, with each iteration of the chosen MH-sampler, $\rho$ will decrease and, in general, so will its estimate $r_A$. We aim to choose a sufficiently small positive threshold, $r_A^*$, to be the point beyond which there is no longer any value in carrying out further MH proposals to rejuvenate the sample, as particles spawned from different progenitors become indistinguishable from each other. Ideally this threshold is as large as is practicably possible, to minimise the number of rejuvenations required and accordingly we test our algorithms with thresholds $r_A^* = 0.1, 0.2, 0.5$.
 In principle, stopping rules that are based, even indirectly, on the number of accepted proposals, can induce bias into the particle-based approximations to the target density. However, here the dependence is sufficiently weak to be of little concern as the stopping time of each chain is dependent on the number of accepted proposals in $n_k - 1$ independent chains as well as itself.

\section{Results from IA SMC algorithms}\label{sec:informative}

Here we focus mainly on the intervention-spanning day $83-120$ interval. In what follows, a hybrid algorithm is adopted, using combinations of three thresholds for $r_A$ with both the continuous and discrete sequential algorithms.

\subsection{Scenario 1: Confirmed case and serological data}\label{sec:scenario1}

\begin{figure}
\centering
\resizebox{0.48\linewidth}{!}{\includegraphics{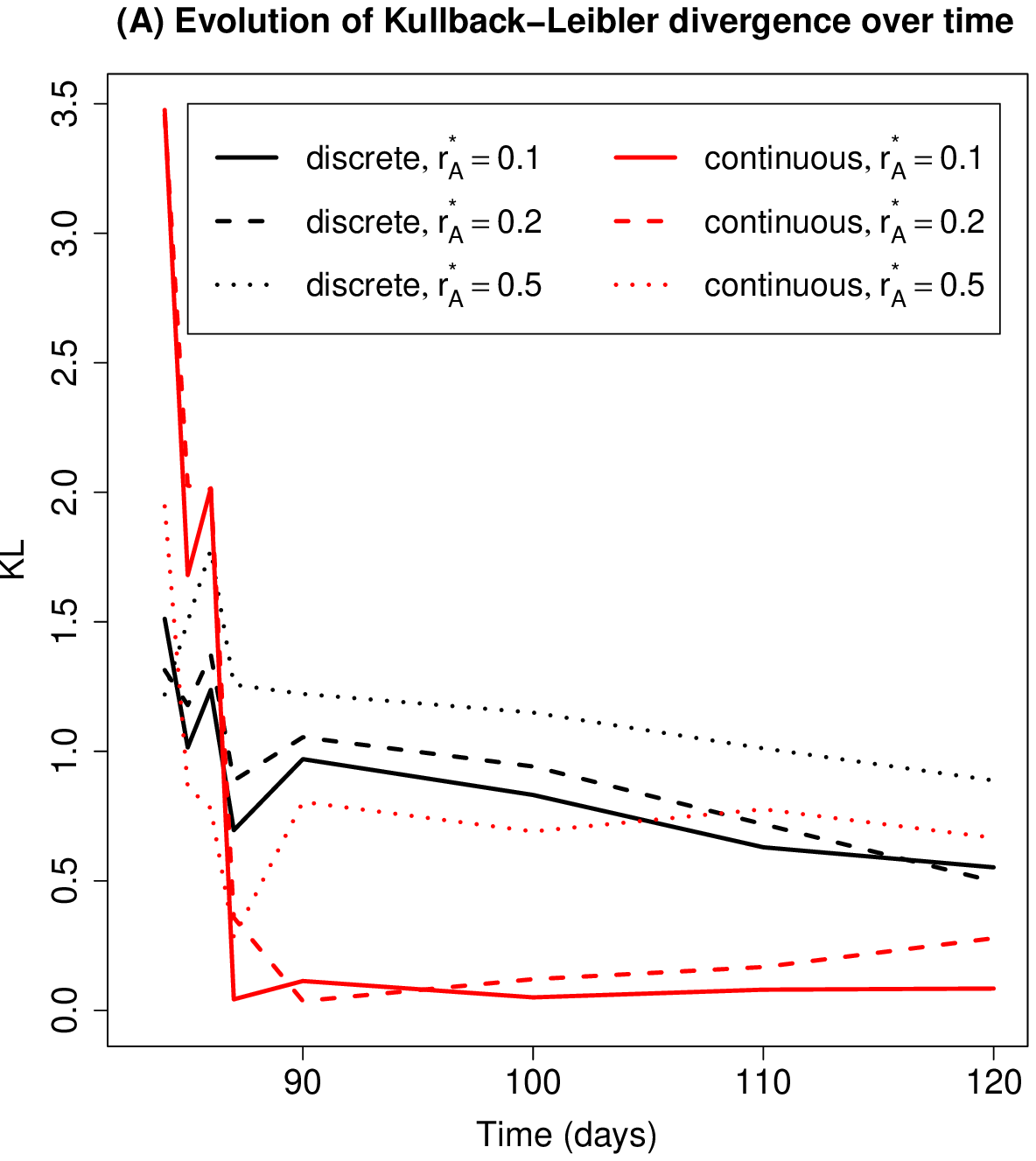}}
\resizebox{0.48\linewidth}{!}{\includegraphics{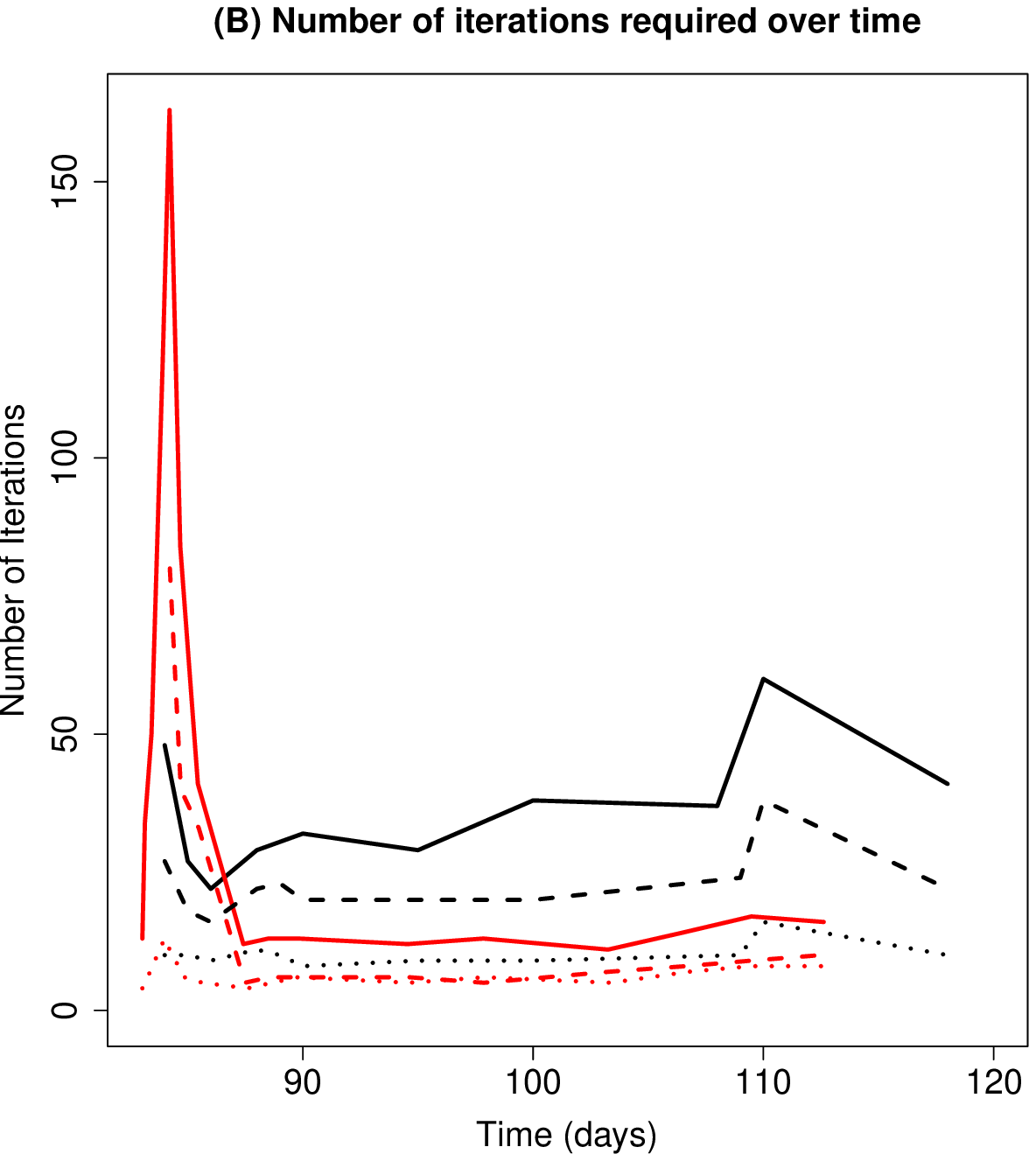}}
\caption{\label{fig:KL6.1} (A) Kullback-Leibler divergence over time;
(B) Number of proposals required at each rejuvenation time by algorithm.}
\end{figure}

MCMC samples were obtained using data up to and including $t_k = 84, 85, 86, 87, 90, 100, 110$ and 120, with Figure \ref{fig:KL6.1} and Table \ref{tbl:inf} summarising the results. 
In Figure \ref{fig:KL6.1}(A), KL discrepancies between $\smcdens{t_k\lvert 83}$ and $\mcmcdens{t_k}$ are plotted over time for each combination of algorithm and threshold. 
To  calibrate these KL divergences, 
a further 40 MCMC chains were obtained at each of these times. The KL divergences between these posterior distributions from the original reference MCMC analysis were then calculated. This formed a distribution of KL values that are typical of MCMC samples from our target distribution. If
$\smcdens{t_k\lvert t_l}$ attains the gold-standard then it should return a KL divergence that could feasibly come from this distribution. Therefore we generate a `KL target' (see Table \ref{tbl:inf}), the $95\%$ quantile of these sampled KL values and diagnose significant difference in the MCMC and SMC-obtained distributions when their KL divergence is larger than this KL target.

Performance of the continuous-time algorithm appears strongly linked to the acceptance rate of the block approximate Gibbs' proposals. This acceptance rate is particularly low (1-2\%) prior to $t_k = 87$, when it undergoes a step change to 15-20\%. 
In contrast, the acceptance rates for the discrete-time algorithm are consistently around 5\% throughout, as seen from the constant number of iterations required over time (Figure \ref{fig:KL6.1}(B)). As a result, from day 87 onwards, far fewer proposals are required in total for the continuous-time algorithm, even if the number of rejuvenation times increases. 

\begin{table*}
\caption{\label{tbl:inf}Performance in scenario 1 of the information-adjusted SMC algorithms over the interval 83-120 days (discrete and continuous) by ICC threshold. 
}
\centering
\begin{tabular*}{\hsize}{@{\extracolsep{\fill}}rrrrrrrr}
\hline
\multicolumn{1}{l}{\bf ICC threshold}&\multicolumn{1}{c}{\bf  0.5}&\multicolumn{1}{c}{\bf  0.2}&{\bf  0.1}&\multicolumn{1}{l}{\bf  ICC threshold}&{\bf  0.5}&{\bf  0.2}&{\bf  0.1}\\
\hline
\multicolumn{4}{l}{{\bf 84 Days} (KL target = 0.732)}&\multicolumn{4}{l}{{\bf 90 Days} (KL target = 0.159)}\\
Continuous&1.95&3.46&3.48&Continuous&0.805&{\bf 0.036}&{\bf 0.113}\\
Discrete&1.22&1.31&1.51&Discrete&1.22&1.05&0.970\\
\multicolumn{4}{l}{{\bf 85 Days} (KL target = 0.135)}&\multicolumn{4}{l}{{\bf 100 Days} (KL target = 0.135)}\\
Continuous&0.862&2.03&1.68&Continuous&0.691&{\bf 0.120}&{\bf 0.050}\\
Discrete&1.50&1.18&1.02&Discrete&1.15&0.942&0.832\\
\multicolumn{4}{l}{{\bf 86 Days} (KL target = 0.365)}&\multicolumn{4}{l}{{\bf 110 Days} (KL target = 0.122)}\\
Continuous&0.780&2.01&2.02&Continuous&0.776&0.167&{\bf 0.080}\\
Discrete&1.78&1.37&1.24&Discrete&1.01&0.719&0.630\\
\multicolumn{4}{l}{{\bf 87 Days} (KL target = 0.276)}&\multicolumn{4}{l}{{\bf 120 Days} (KL target 0.119)}\\
Continuous&0.282&0.358&{\bf 0.043}&Continuous&0.666&0.278&{\bf 0.084}\\
Discrete&1.26&0.887&0.696&Discrete&0.888&0.498&0.552\\
\hline
\end{tabular*}
\end{table*}

\subsection{Scenario 2: Primary care consulation and serology data}\label{sec:last.analysis}

\begin{figure}[!t]
\resizebox{0.3243825\linewidth}{!}{\includegraphics{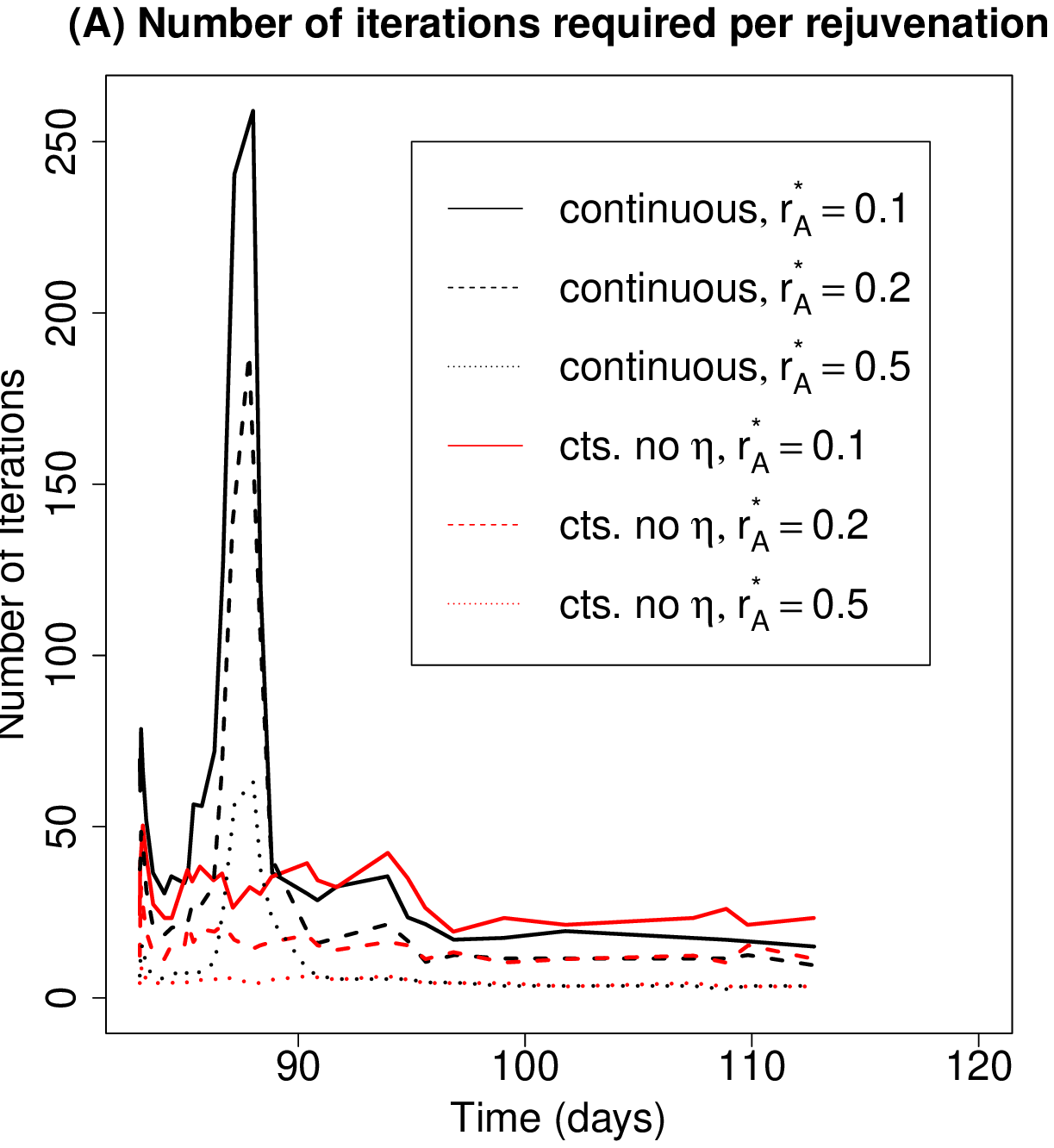}}\resizebox{0.3243825\linewidth}{!}{\includegraphics{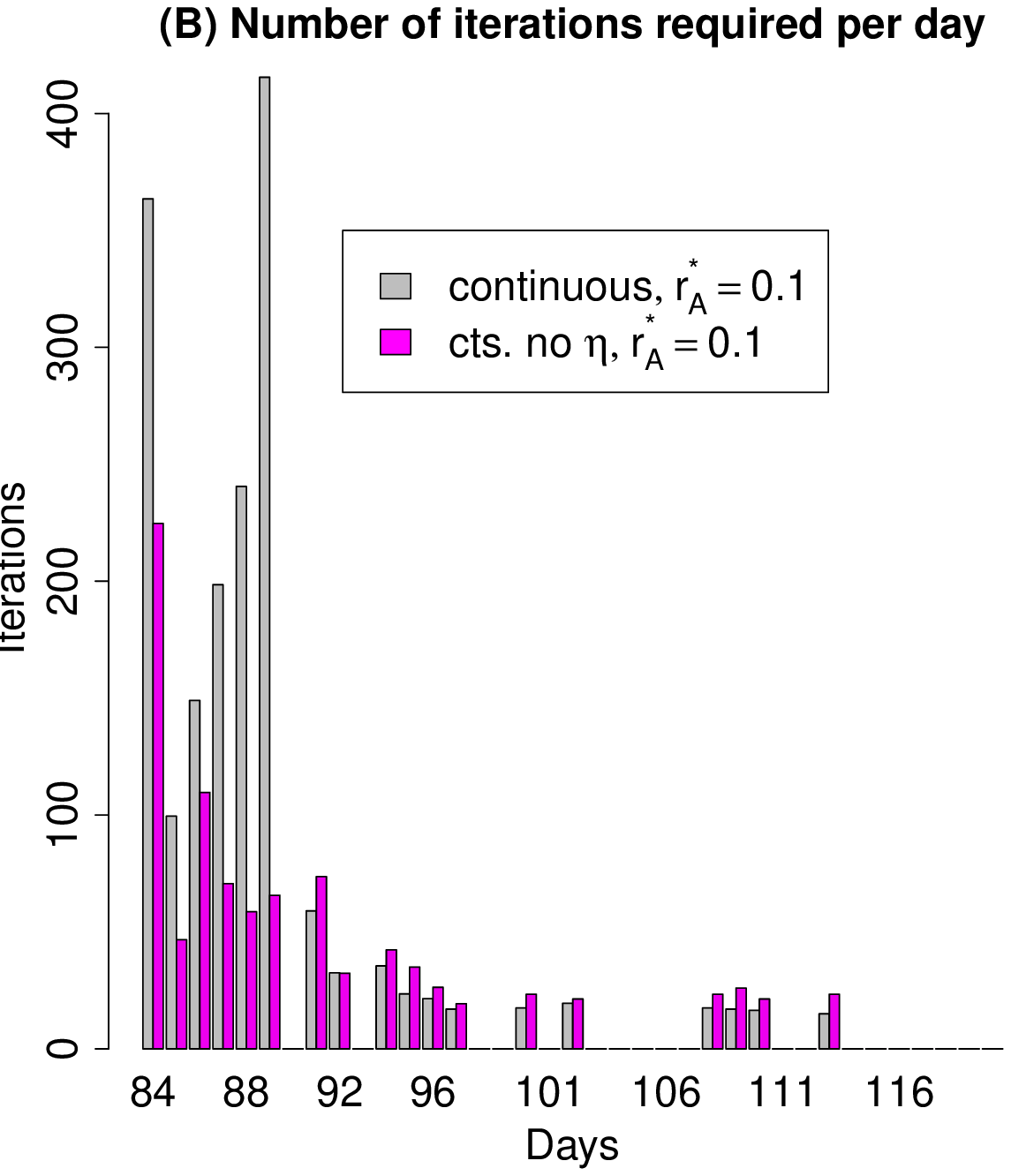}}\resizebox{0.3243825\linewidth}{!}{\includegraphics{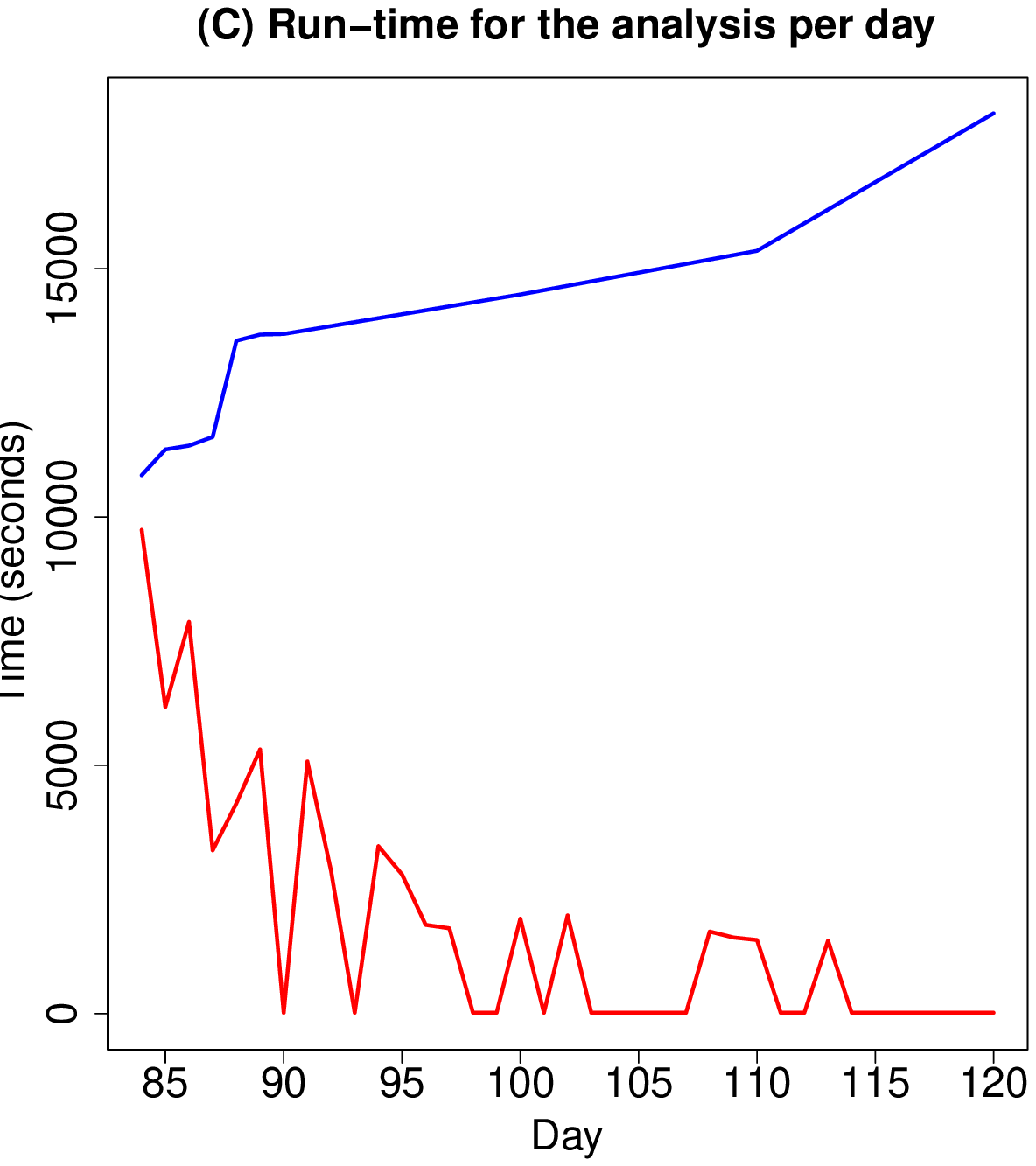}}
\caption{\label{fig:eta2}(A) Number of MH-steps required by the continuous-time SMC algorithms per rejuvenation over time; (B) Total number of MH-steps required by the continuous-time SMC algorithms per time interval; (C) The computation time required for model runs on each day using MCMC (blue line) and SMC (red line).}
\end{figure}

Focusing on the better-performing continuous-time IA algorithm, similar performance to Scenario 1 can be observed (Table \ref{tbl:inf2.alt}).
The algorithm again suffers from acceptance rates for the approximate-Gibbs' proposals, which, though initially adequate, fall to 0.3\% on day 89, illustrated by a peak of over 250 proposals per rejuvenation and over 400 proposals per day in Figures \ref{fig:eta2}(A) and (B) respectively.  
This low rate is driven by the highly non-Gaussian distribution for the dispersion parameter $\eta_2$, which  has an unbounded gamma prior and is not well identified from the data. To improve acceptance rates a `cts.reduced' scheme is devised in which the dispersion parameters are omitted from the block approximate-Gibbs updates and proposed separately. In terms of the resulting KL divergences, there is no significant drop in performance between the continuous to the `cts. reduced' algorithm (Table \ref{tbl:inf2.alt}). 
The `cts. reduced' proposal scheme requires far fewer iterations of the Metropolis-Hastings algorithm over the interval 84-90 days, maintaining acceptance rates of about 10\% over this period. On day 90, the `cts.reduced' scheme does give an anomalously high KL value (1.42). Closer inspection found this to be the result of three particles with extremely small values for $\eta$. With these three particles removed, the KL divergence falls to 0.086. Over time, as the target distribution converges to a multivariate normal distribution, the number of moves required for both methods equalise 
and the benefit of the `cts. reduced' proposal scheme vanishes (Figure \ref{fig:eta2}B).

\begin{table*}
\caption{\label{tbl:inf2.alt}Performance in scenario 2 of the information-adjusted SMC algorithm over the interval 83-120 days in continuous time where the algorithms differ in the inclusion of the $\eta$ parameters in the block
proposals. Parameter $\vec{\beta}^B$ is omitted from the KL calculations.}
\centering
\begin{tabular*}{\hsize}{@{\extracolsep{\fill}}rrrrrrrr}
\hline
\multicolumn{1}{l}{\bf  ICC threshold}&{\bf  0.5}&{\bf  0.2}&{\bf  0.1}&\multicolumn{1}{l}{\bf  ICC threshold}&{\bf  0.5}&{\bf  0.2}&{\bf  0.1}\\
\hline
\multicolumn{4}{l}{{\bf 84 Days} (KL target = 6.06)}&\multicolumn{4}{l}{{\bf 90 Days} (KL target = 0.120)}\\
Continuous&{\bf 2.92}&{\bf 2.87}&{\bf 2.83}&Continuous&1.80&0.35&{\bf 0.066}\\
Cts. Reduced&{\bf 2.97}&{\bf 2.85}&{\bf 2.86}&Cts. Reduced&2.10&{\bf 0.093}&1.42\\
\multicolumn{4}{l}{{\bf 85 Days} (KL target = 1.90)}&\multicolumn{4}{l}{{\bf 100 Days} (KL target = 0.182)}\\
Continuous&3.05&3.00&2.98&Continuous&{\bf 0.157}&{\bf 0.102}&{\bf 0.089}\\
Cts. Reduced&3.06&2.97&2.98&Cts. Reduced&{\bf  0.107}&{\bf 0.084}&{\bf 0.070}\\
\multicolumn{4}{l}{{\bf 86 Days} (KL target = 1.94)}&\multicolumn{4}{l}{{\bf 110 Days} (KL target = 0.0936)}\\
Continuous&3.28&3.24&3.25&Continuous&0.159&{\bf 0.077}&0.111\\
Cts. Reduced&3.27&3.22&3.26&Cts. Reduced&0.197&{\bf 0.037}&{\bf 0.035}\\
\multicolumn{4}{l}{{\bf 87 Days} (KL target = 5.44)}&\multicolumn{4}{l}{{\bf 120 Days} (KL target = 0.101)}\\
Continuous&{\bf 2.54}&{\bf 2.45}&{\bf 2.42}&Continuous&0.136&{\bf 0.044}&{\bf 0.071}\\
Cts. Reduced&{\bf 2.51}&{\bf 2.48}&{\bf 2.44}&Cts. Reduced&{\bf 0.100}&{\bf 0.042}&{\bf 0.055}\\
\hline
\end{tabular*}
\end{table*}

\begin{figure}
\centering
\hspace{0.09\linewidth}\textbf{\textsf{SMC}}\hspace{0.18\linewidth}\textbf{\textsf{MCMC}}\hspace{0.205\linewidth}\textbf{\textsf{SMC}}\hspace{0.18\linewidth}\textbf{\textsf{MCMC}}\hspace{0.07\linewidth}
\begin{tabular}{c|c}
\resizebox{0.499\linewidth}{!}{\includegraphics{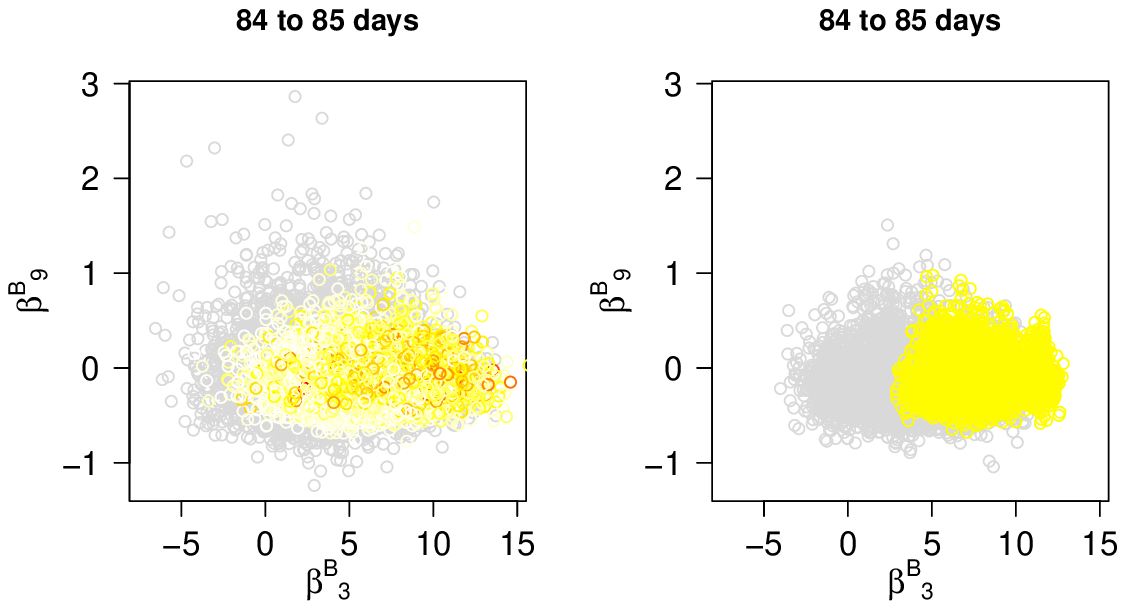}}&\resizebox{0.499\linewidth}{!}{\includegraphics{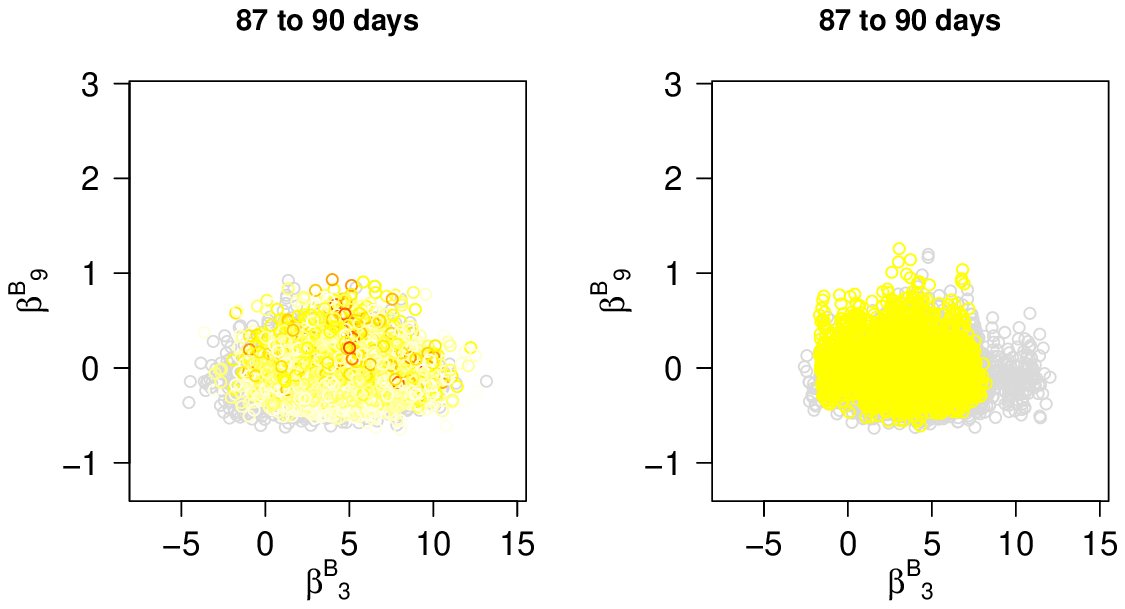}}\\
\resizebox{0.499\linewidth}{!}{\includegraphics{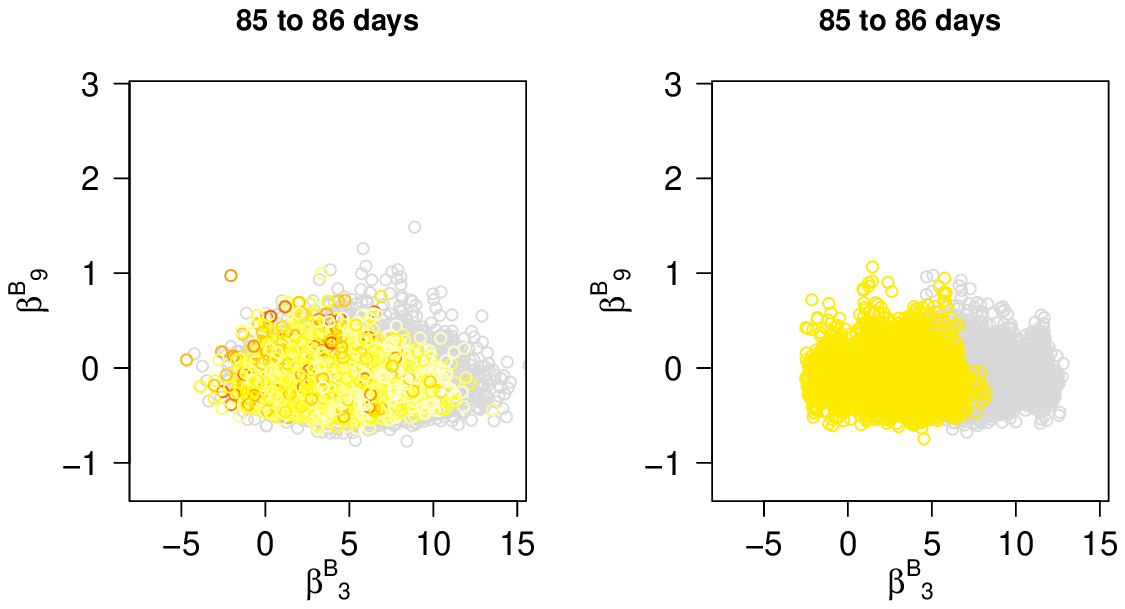}}&\resizebox{0.499\linewidth}{!}{\includegraphics{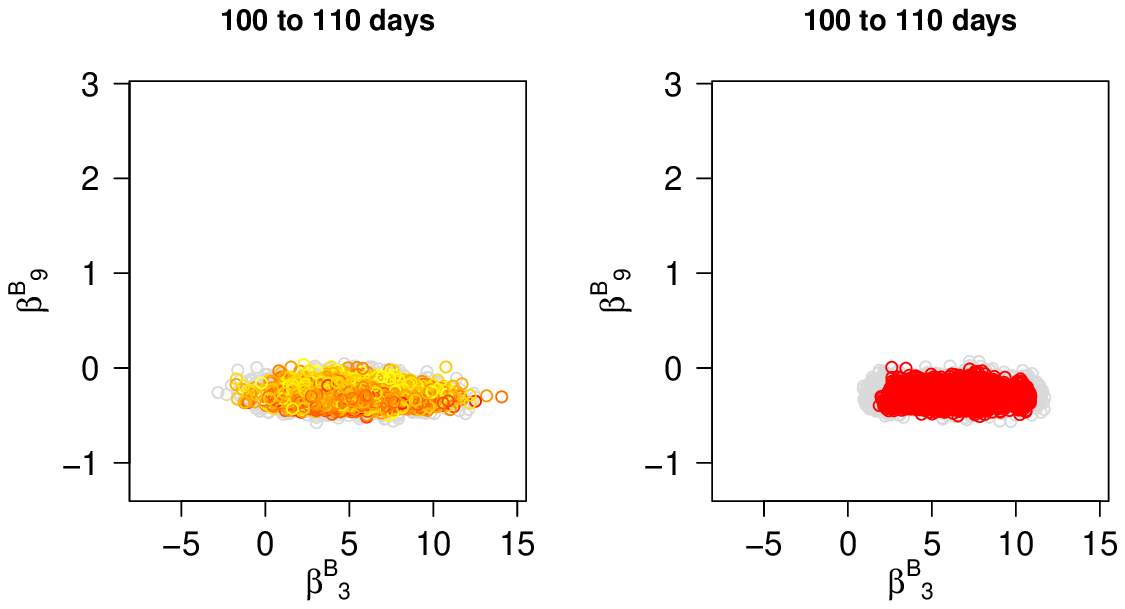}}\\
\end{tabular}
\caption{\label{fig:Bij.evo}The evolution over time of the marginal joint posterior for two components of the parameter vector $\beta^B$. Comparison between SMC-obtained and MCMC-obtained posterior distributions. Grey points indicate the distribution at the start of the interval.}
\end{figure}

The scatter plots of Figure \ref{fig:Bij.evo} give a sequence (over time) of marginal posterior distributions for two parameters, $(\beta^B_3, \beta^B_9)$, of the background consultation rate model, obtained from the `cts.reduced' SMC scheme and MCMC. These parameters are only weakly identifiable in the immediate period after $t_k = 83$ and a clear discrepancy between the MCMC- and the SMC-obtained posterior scatters emerges. The SMC distributions, being based on many short MCMC chains, cover the full posterior distribution adequately at each $t_k$. The MCMC has difficulty mixing, at $t_k = 85, 86$ in particular, resulting in scatters concentrated in a sub-region of the full marginal support. 

Not only does SMC offer an improvement in terms of posterior coverage in the presence of partial identifiability, but its daily implementation is also faster, as shown in Figure \ref{fig:eta2}(C).
The runtime for SMC decreases almost linearly with increased parallelisation and so the particles (and hence the parallel MCMC chains) are distributed across 255 Intel(R) Xeon(R) CPU E5-262 2.0GHz processors on a high-performance computing cluster. This represents modest parallelisation compared to what might be used in a real pandemic. Figure \ref{fig:eta2} shows that, not only is SMC more computationally efficient on day 84, the day requiring the most MH-updates to rejuvenate the sample, but the run-times decrease over time, in contrast to the increasing MCMC run-times as more data have to be analysed. On days where the sample does not require rejuvenation, run-times are negligible.

\section{Discussion}\label{sec:discussion}

This paper addresses the substantive problem of online tracking of an emergent epidemic, assimilating multiple sources of information through the development of an information-adjusted SMC algorithm. When incoming data follow a stable pattern, this process can be automated using standard SMC algorithms, confirming current knowledge \citep[\textit{e.g.}][]{DukLP12,OngCCLLLTG10}. However, in the likely presence of interventions or any other event that may provide a system shock, it is necessary to adapt the algorithm appropriately.

Using a simulated epidemic where a public health intervention provides a sudden change to the pattern of case reporting, we have constructed a more robust SMC algorithm by tailoring: 
1) the choice of rejuvenation times through tempering; 2) the  choice of the MH-kernel by combining local random walk and Gibbs proposals;
3) a stopping rule for the MH steps based on intra-class correlations to minimise the number of iterations within each rejuvenation.

The end result is an algorithm that is a hybrid of particle filter and population MCMC \citep{Gey91, LiaW01, JasSH07}; is robust to possible shocks; improves over the plain-vanilla MCMC in terms of run-times needed to derive accurate inference;
and can automatically provide all the distributions needed for posterior predictive measures of model adequacy.

\subsection{Benefits of SMC}\label{sec:benefits}

\paragraph{Model Run Times}
From a computational point of view, the SMC algorithm is faster than the plain vanilla MCMC as it is highly parallellisable. However, this may be an unfair comparison as 
we could have considered more sophisticated MCMC algorithms, as exemplified in an epidemic context by \cite{JewKCR09}. The use of differential geometric MCMC \citep{GirC11}, non-reversible MCMC \citep{BieFR16} or MCMC using parallelisation \citep{BanGLRarX} could improve run times. However, as MCMC steps are the main computational overhead of the SMC algorithm, any improvements to the MCMC algorithm's efficiency may also improve the SMC. As target posteriors attain asymptotic normality it should be progressively easier for SMC to move between distributions over time, as can be seen in Figure \ref{fig:eta2}(C) where the daily running time decreases as data accumulate. For any MCMC algorithm, the opposite will be generally true.

\paragraph{Predictive Model Assessment}

\begin{figure}
\centering
\resizebox{0.328\linewidth}{!}{\includegraphics{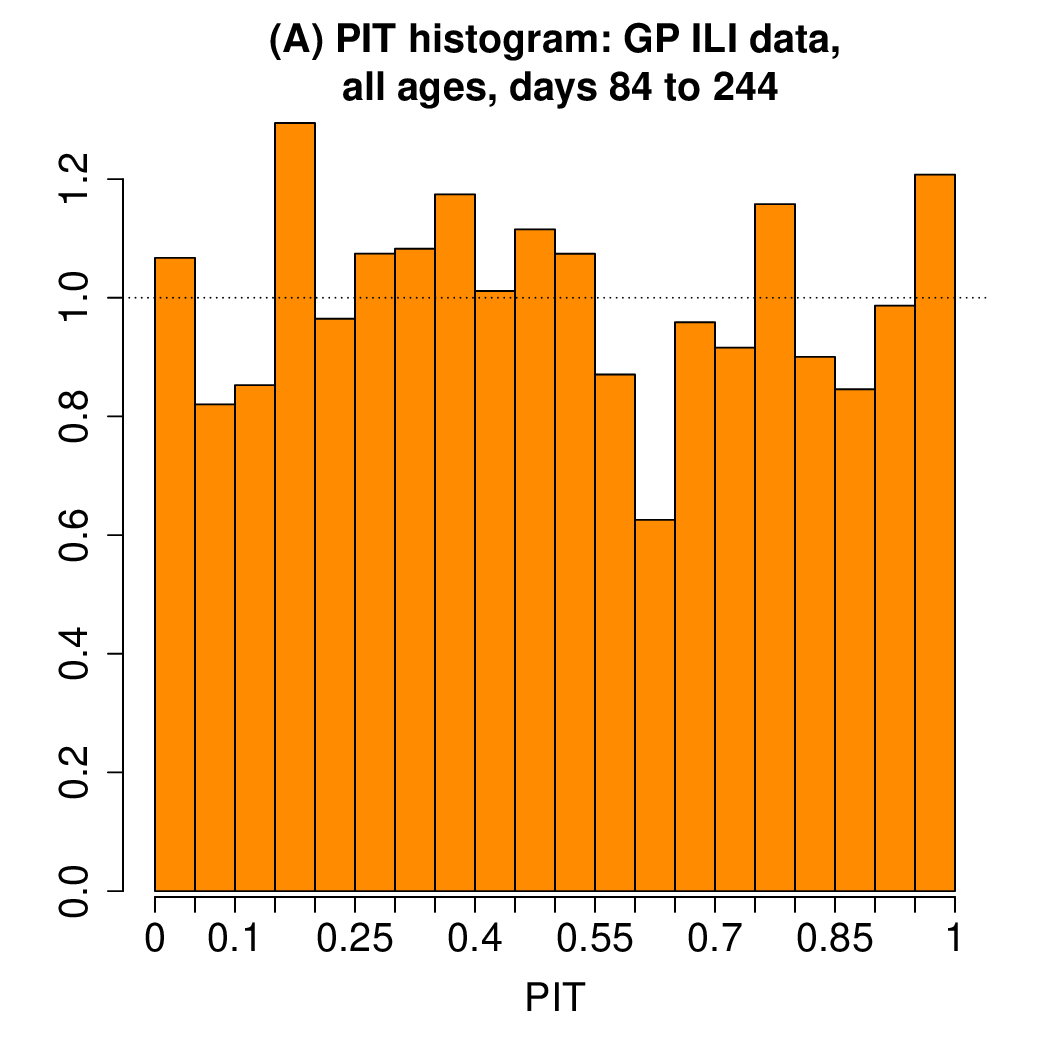}}
\resizebox{0.328\linewidth}{!}{\includegraphics{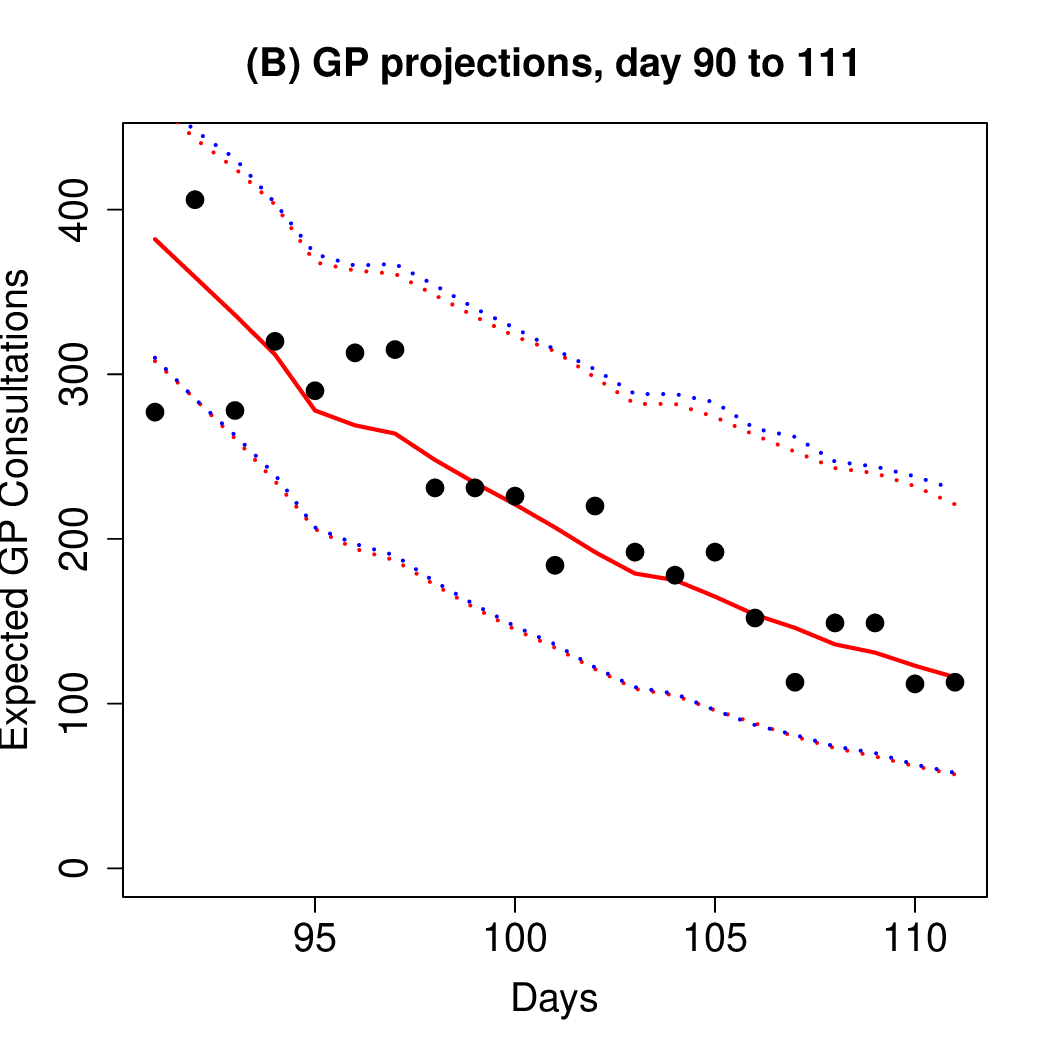}}\resizebox{0.328\linewidth}{!}{\includegraphics{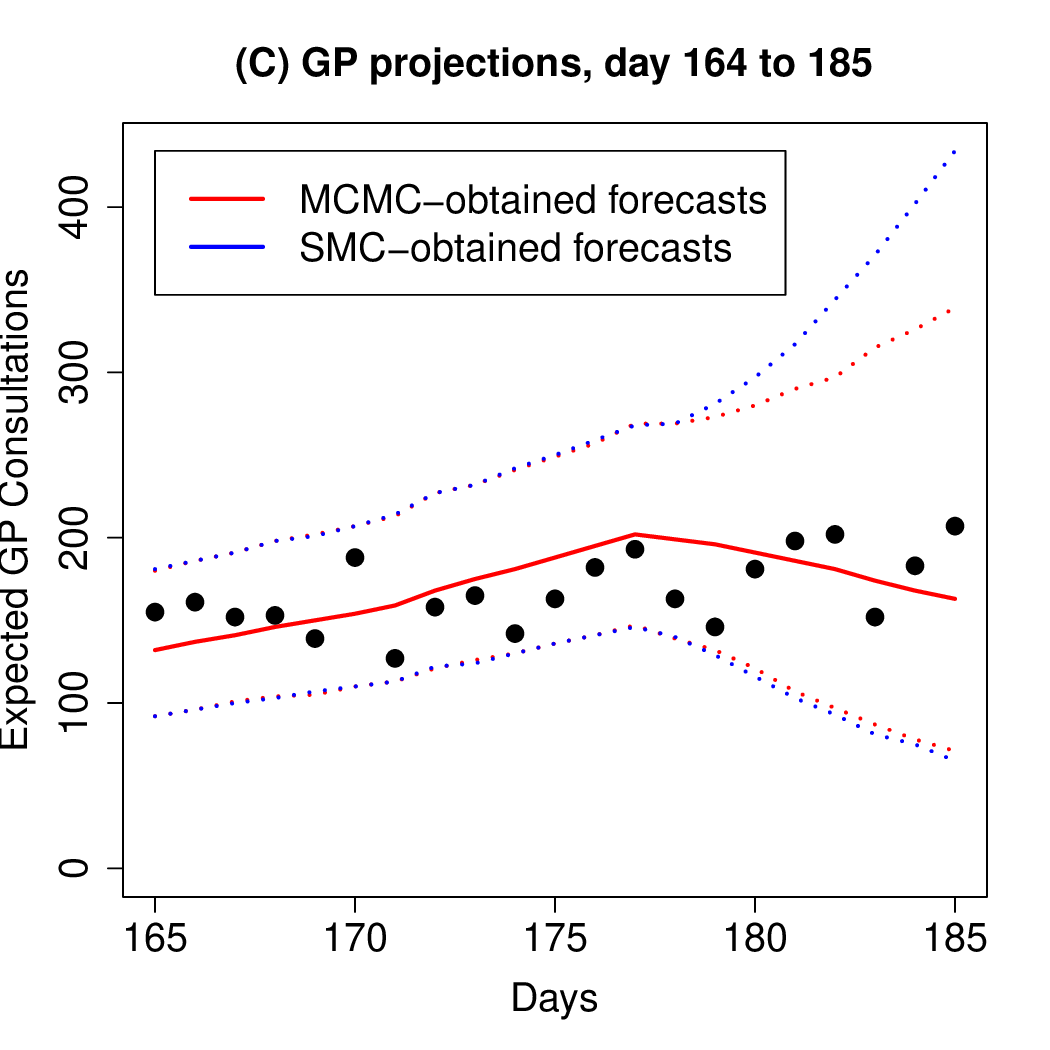}}
\caption{\label{fig:forecasts}(A) PIT histograms for the one-step ahead predictions of GP ILI consultation data, calculated over.  162$\times$7 time and strata combinations. (B) and (C) Comparison of the observed GP data with posterior predictive distributions obtained using the SMC and MCMC algorithms at day 90 and 164 respectively. Solid lines give posterior medians of the distributions and the dotted lines give 95\% credible intervals for the data.}
\end{figure}

A fundamental goal of real-time modelling is to provide online epidemic forecasts with an appropriate quantification of the associated uncertainty. The real time assessment of the predictive adequacy of a model becomes, therefore, key and can be carried out through the evaluation of one-step ahead forecasts based on posterior predictive distributions $p(\data{k+1}\lvert\data{k})$ \citep{Daw84}. Such assessments can be made informally through, for example, probability integral transform (PIT) histograms \citep{CzaGH09}. In the example of Section \ref{sec:last.analysis}, Figure \ref{fig:forecasts}(A) shows the PIT histogram for one-step ahead prediction of primary care consultations for all age groups for successive analyses in the range 84-245 days. A good predictive system would give a uniform histogram and though the histogram here is not entirely uniform, it shows no consistent under or over-estimation nor any clear signs of over-dispersion.

More formally, proper scoring rules \citep{GneR07} can be used to assess the quality of forecasts, including through formal tests of prediction adequacy \citep{SeiD93}. Many different scoring rules exist, but to illustrate a benefit of an SMC algorithm consider 
the logarithmic score, defined, for a predictive distribution $p(\cdot)$ and a subsequently realised observation $y$, to be:
\begin{equation*}
s_{\textrm{log}}(P, y) = -\log (p(y)).
\end{equation*}
Under an SMC scheme, for one step ahead forecasts, these are: 
\begin{align*}
s_{\textrm{log}}(P, y) &= -\log (p(\data{k + 1}\lvert\data{1:k}))\\
&= -\log \left(\int_{\Theta}\pi\left(\vtheta\lvert\data{1:k}\right)p(\data{k+1}\lvert\vtheta)d\vtheta\right)\\
&\approx -\log \left(\sum \wt{k}{j}\lik{k + 1}{\vtheta_{k}^{(j)}} / \sum \wt{k}{j}\right)\\
&= \log \left(\frac{\sum_j \wt{k}{j}}{\sum_j \tilde\omega_{k + 1}^{(j)}}\right).
\end{align*}
Weights $\wt{k}{j}$ and $\tilde\omega_{k + 1}^{(j)}$ are routinely calculated as part of the SMC algorithm in Section \ref{sec:alg} (equation \eqref{eqn:alg.reweight}), whereas additional computation is required if the posterior is derived using MCMC. 
If the MCMC analyses are not carried out with every new batch of data, then these are not readily available. 
For further details on the calculation and interpretation of these posterior predictive methods see Section E of the online appendix.

Figures \ref{fig:forecasts}(B) and (C) present longer-term (3 week) forecasts for the consultation data obtained via both SMC and MCMC from days 90 and 164 onwards. Whereas in (B) the forecasts are close enough to be identical, there is a divergence in the predictive intervals from $t_k =178$ onwards, a change-point in the model for the background ILI rate. 

\paragraph{Identifiability}
As observed in Section \ref{sec:last.analysis}, the SMC algorithm is better at exploring the full posterior distribution in the presence of parameter non-identifiability around changepoints. 
The background ILI rate is modelled using a piecewise log-linear curve (Equation (1) in the online appendix) 
with linear interpolation giving the value of the curve at intervening points.
This results in log-consultation rates in the three days following the changepoint on day 83 that include the respective sums (neglecting the age effects)
$\mu + \alpha_{84}$,
$\mu + 0.98\alpha_{84} + 0.02\alpha_{128}$,
$\mu + 0.96\alpha_{84} + 0.04\alpha_{128}$.
This makes parameters $\mu$ and $\alpha_{84}$ only weakly identifiable over this period, inducing convergence problems for MCMC (see Figure \ref{fig:Bij.evo}). Further evidence for this is given in Figure \ref{fig:forecasts}(C) by the divergence of the prediction intervals at breakpoint $t_k = 178$ and in Table D3 in the online appendix, where the KL calculations of Table \ref{tbl:inf2.alt} are repeated but with background parameters included. The marked increases in the KL targets from day 90 onwards is a result of significant discrepancy between the MCMC chains.
\cite{JasSDT11} claim that, for their example, SMC may well be superior to MCMC and this is one case where this is certainly true. The population MCMC carried out in the rejuvenation stage achieves good coverage of the sample space, without the individual chains having to do likewise. Reparameterisation may improve the MCMC, but this would also be of benefit to the SMC rejuvenation steps.

\paragraph{Early Warning}
Changepoints that lead to the lack of identifiability discussed above may coincide with public health interventions. In this paper, it is assumed that such times are known and we have been concerned with the adaptation of inferential procedures to ensure that they can be operated in a semi-automatic fashion at such times.

In general such changepoints will need to be detected in real time and may be indicative of a change in the underlying epidemic dynamics or in healthcare-seeking behaviours, both of which are of great interest to healthcare managers. A sudden drop in the ESS can raise a flag that the model is no longer suitable and may require modification. Both \cite{WhiJG11} and \cite{NemFM14} discuss automated approaches for the sequential detection of changepoints. However, when considering a complex mechanistic epidemic model a more fundamental adaptation may be required. Sequential application of MCMC as data arrive over time would not automatically detect this without carrying out a series of exhaustive post-hoc diagnostic checks.

\subsection{Final considerations}
In answer to the question initially posed, we have provided a recipe for online tracking of an emergent epidemic using imperfect data from multiple sources. We have discussed many of the challenges to efficient inference, with particular focus on scenarios where the available information is rapidly evolving and is subject to sudden shocks. Throughout we have inevitably made pragmatic choices and alternative strategies could have been adopted.
The choice of the MH-kernels used for rejuvenation is an example. There are many options to tweak the performance of the ``vanilla'' kernels presented here, including: simply scaling the covariance matrix in the approximate-Gibbs moves \citep{Wes93}; treating the composite proposals of Section \ref{sec:kernels} as a single mixture \citep{KanBJ14}; using recent developments in kernel SMC methods to design local covariance matrices \citep{SchSPS17}; and incorporating an adaptive scheme to select an optimal SMC kernel and any tuning parameters \citep{FeaT13}. Equally, we could have adopted 
multivariate analogues for the intra-class correlation coefficient \citep[e.g.][]{Ahr76, KonKR91} to define a rejuvenation stopping rule; or we could have opted for a particle set expansion by increasing $n_k$ as a possible alternative to running long MCMC chains for each particle when new parameters are introduced in the model, for example through a shock.

We have shown above that the benefits of SMC for online inference extend beyond computational efficiency. It is not claimed, however, that SMC is beneficial when inference is carred out offline, using the full available data. Over the course of any outbreak, the richness of data may grow, interventions may occur and models of increased complexity may be needed. It is therefore important to retain the capacity to fit new models efficiently. 
Methods such as Hamiltonian MCMC \citep{GirC11}, likelihood-tempered SMC algorithms \citep{KanBJ14}, emulation \citep{FarBCD14}, variational \citep{BleKM17} and Kalman-filtering approaches \citep{ShaK12} represent potential alternatives to achieve this.

We have focused on an epidemic scenario that has the potential to arise in the UK. Nevertheless, our approach addresses modelling concerns common globally \citep[e.g.][]{WuCLIHTCLLLR10, ShuLLA16, TebBWDV15} and can form a flexible basis for real-time modelling strategies elsewhere. Real-time modelling is, however, more than just a computational problem. It does require the timely availability of relevant data,  a sound understanding of any likely biases, and effective interaction with experts. In any country, only interdisciplinary collaboration between statisticians, epidemiologists and database managers can turn cutting edge methodology into a critical support tool for public health policy. 

\Urlmuskip=0mu plus 2mu\relax


\end{document}